\documentclass[%
 amsmath,amssymb,
 aps, physrev,
]{revtex4-2}
\usepackage[T1]{fontenc}
\usepackage{mathtools}
\usepackage{graphicx}
\usepackage{dcolumn}
\usepackage{bm}
\usepackage{braket}
\usepackage{xcolor}
\usepackage[normalem]{ulem}

\begin{document}

\preprint{}

\title{\textbf{Demonstration of a photonic time-frequency Fourier transform and temporal double slit using atomic quantum memory} 
}%

\author{Ankit Papneja}
\author{Jesse Everett}
\author{Cameron Trainor}
\author{Aaron D. Tranter}
\author{Ben C. Buchler}
\email{ben.buchler@anu.edu.au}

\affiliation{%
Research School of Physics\\
Australian National University\\
Acton 2601, Australia
}

\date{\today}

\begin{abstract}
A quantum memory for light is expected to play a crucial role in quantum communication protocols and distributed quantum computing.  In addition to storage and buffering, a quantum memory can be used for manipulations of stored states to allow more complex quantum network operations. In this work, we demonstrate an in-memory Fourier transform using a combination of two well-established quantum memory protocols: Gradient Echo Memory and Electromagnetically Induced Transparency. Our experiment is realised using an ensemble of rubidium atoms that are laser cooled in an elongated magneto-optic trap to maximise optical depth.  The results of our time-frequency Fourier transform can be understood as a temporal double slit.  We show that the interference between time-separated pulses depends on the relative phase and time between the pulses of light.  The use of a quantum memory enables us to illuminate exactly where and how interference occurs between time separated pulses. Time-frequency Fourier manipulation is a well established technique in classical optical systems. Our combination of Fourier manipulation and quantum-compatible memory could be used to bring similar capability to quantum optical systems. 
\end{abstract}

\maketitle

\section{Introduction}
Manipulation of the temporal and spectral properties of light is ubiquitous in optical signal processing. A set of techniques for spectro-temporal transformations has emerged based on dispersion and frequency manipulation
\cite{vanhoweUltrafastOpticalSignal2006,tsangPropagationTemporalEntanglement2006} as a means of adjusting
single photon spectra \cite{donohueSpectrallyEngineeringPhotonic2016,matsudaDeterministicReshapingSinglephoton2016,karpinskiBandwidthManipulationQuantum2017,luQuantumInterferenceCorrelation2018,francesconiEngineeringTwophotonWavefunction2020,sosnickiAperiodicElectroopticTime2020}, temporal mode sorting and characterisation \cite{joshiPicosecondresolutionSinglephotonTime2022, lipkaSuperresolutionUltrafastPulses2024}, and for processing optical quantum signals inside a memory \cite{mazelanikTemporalImagingUltranarrowband2020a, mazelanikOpticaldomainSpectralSuperresolution2022a,nieweltExperimentalImplementationOptical2023}. A time-frequency Fourier transform (TFFT) is one application of these techniques, partly inspired by parallels between spectro-temporal manipulation for optical signal processing and the spatial diffraction and propagation of light for optical imaging \cite{kolnerSpacetimeDualityTheory1994}. These methods do not call on amplification or lossy filtering of the signal, and are therefore relevant for quantum optical signal processing \cite{karpinskiControlMeasurementQuantum2021}.

Quantum memories for light are expected to play a key role in distribution of entanglement \cite{Lei:23}, which is an essential part of long range communication systems \cite{RevModPhys.95.045006} and distributed quantum computing \cite{BARRAL2025100747}. 
Here we present a method for a TFFT based on a combination of two quantum memory protocols. We combine the Gradient Echo Memory (GEM) with Electromagnetically-Induced Transparency (EIT), and use the different spatial encodings in the two memories to realise an intra-memory TFFT.

The atom ensemble EIT memory is based on the creation of ultra-slow light. When a freely propagating optical pulse is compressed into an EIT polariton, the signal is spatially compressed into the atomic ensemble, thus encoding the \textit{time} information in the longitudinal position of the coherent atomic excitation \cite{lvovskyOpticalQuantumMemory2009a, harrisElectromagneticallyInducedTransparency1997}. Numerous works have demonstrated the use of EIT to enable effective quantum memory with high efficiency \cite{hsiaoHighlyEfficientCoherent2018a} and preservation of quantum states \cite{vernaz-grisHighlyefficientQuantumMemory2018}. In GEM, a longitudinal frequency gradient in the atomic absorption spectrum encodes the \textit{frequency} information of incoming light in the longitudinal position of the atomic coherence used for the memory \cite{moiseevEfficiencyFidelityPhotonecho2008, hetetMultimodalPropertiesDynamics2008, campbellEchoBasedQuantumMemory2016}. This protocol has also demonstrated high efficiency \cite{choHighlyEfficientOptical2016} and compatibility with quantum states \cite{leungHighlyEfficientStorage2024}.

The key to the current work is the combination of these techniques in a single system. We show that it is possible to store light with the GEM protocol and retrieve it with EIT, swapping the encoding and decoding domains and thereby applying a TFFT. We demonstrate this transform in a cold atomic ensemble memory with a `temporal double slit' \cite{tiroleDoubleslitTimeDiffraction2023, hongDelayedchoiceQuantumErasure2023,lindnerAttosecondDoubleSlitExperiment2005, richterStreakingTemporalDoubleSlit2015,PhysRevLett.89.173001,PhysRevLett.77.4}. In a spatial double slit, a fringe pattern is observed in the far-field or Fourier plane of an imaging lens. The application of a TFFT to a temporal double slit leads to an analogous fringe pattern in the temporal envelope of the output of our atomic memory.

Our experiment is by no means the first to be done in the optical domain. Previous experiments have shown both first- \cite{tiroleDoubleslitTimeDiffraction2023} and second-order \cite{hongDelayedchoiceQuantumErasure2023} interference. Nevertheless, our work has some distinguishing features that allow us to show exactly where and how interference between temporally separated pulses occurs, as we will discuss in the context of our experimental results.

\section{Theoretical description of GEM-EIT}
\begin{figure}[h!]
    \centering
    \includegraphics[width=0.8\linewidth]{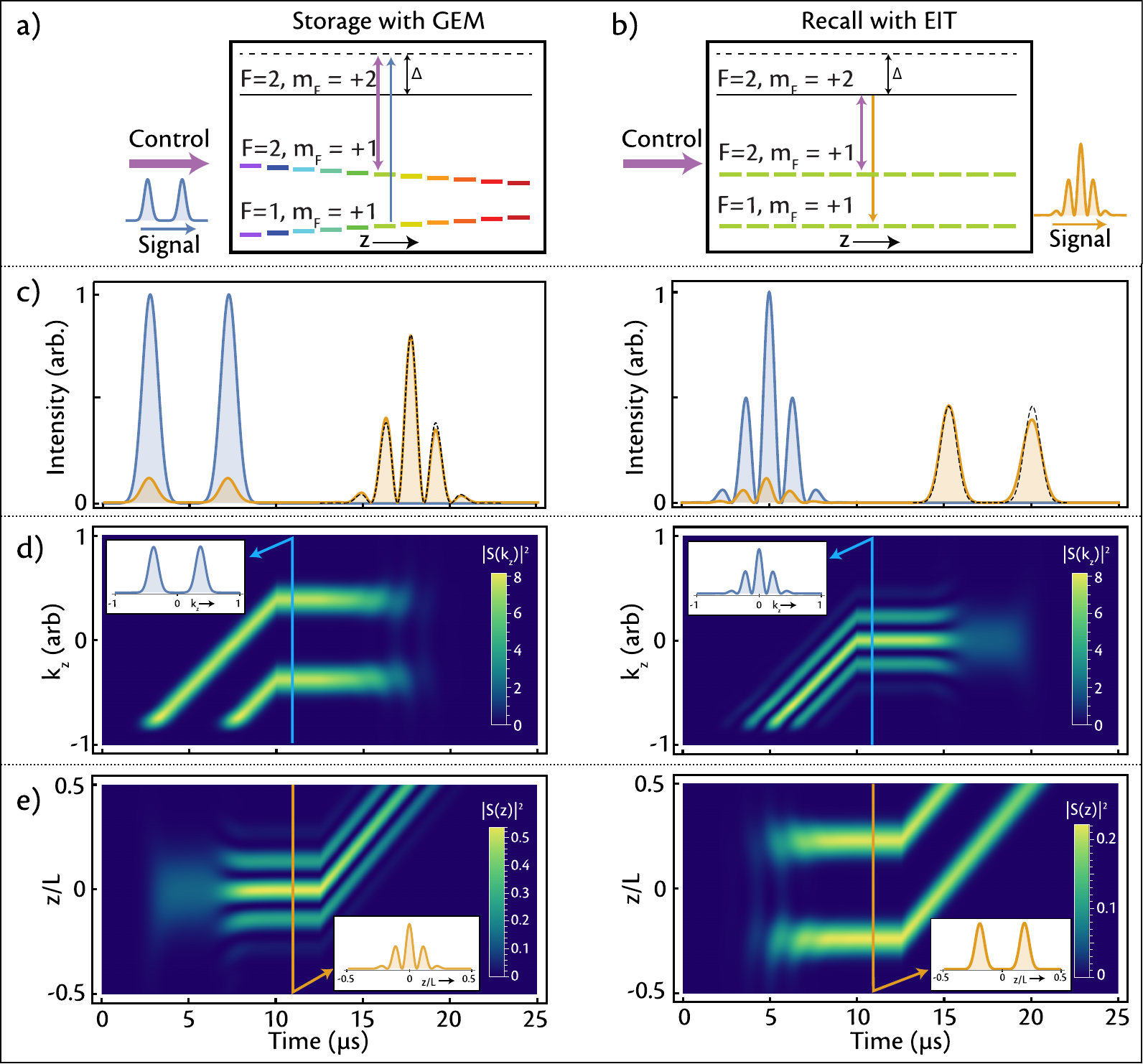}
    \caption{
a) Level scheme for GEM showing the spatial gradient in the two-photon absorption and a large detuning from the excited state. b) Level scheme for EIT with no spatial gradient and optical fields resonant with the excited state. Not shown in these diagrams is the two-photon detuning, $\delta$, see Methods for details.
Simulation results for intra-memory TFFT are shown in (c-e).  The left column shows results for a double pulse input, while the right column shows results for a modulated Gaussian input.  c) Input (blue) and output (orange) envelopes. A scaled Fourier transform (dashed) of the input envelope is placed over the output for comparison. d) Momentum distribution of the spinwaves, showing that GEM writes the input pulse envelopes into the momentum space of the spinwave. The insets show cross sections in momentum space after GEM storage into the spinwave is complete. e) Spatial distribution of the spinwaves, with EIT reading the spatial component as the time envelope of the output signal. The insets show cross sections in position space after GEM storage into the spinwave is complete.
}
\label{fig:concept}
\end{figure}

Although EIT and GEM have distinct properties, they are both very well modelled by the optical Bloch equations (OBE) for three-level $\Lambda$ atomic ensembles \cite{gorshkovPhotonStorageType2007a}.
To simulate EIT or GEM, the input to the OBE model is varied in two key ways that are summarised in Figs.~\ref{fig:concept} (a) and (b). Firstly, the control beam detuning varies between the methods. In GEM, a far-detuned optical control field stimulates the Raman absorption and re-emission of the signal.
In EIT, on the other hand, the control field is resonant and the intensity of the control field regulates the group velocity of signal light within the EIT window, allowing for stopped light as the control field is turned off, and recall as it is turned back on.
The second key difference is that in GEM, a longitudinal atomic frequency gradient is applied to the atoms so that the frequency of Raman absorption varies linearly along the length of the memory. 
To retrieve the stored light with GEM, this gradient is reversed in sign leading to a rephasing of the atomic coherence and re-emission in the presence of the control field.

In EIT, the control field intensity regulates the memory bandwidth, storage and recall. In GEM, bandwidth is determined by the atomic frequency gradient, while storage and recall rely on the combination of control field intensity and the rephasing that is regulated by the atomic frequency gradient. 
The control field detuning, intensity and atomic frequency gradient can be changed when the excited state coherence is negligible without invalidating the OBE. Similar approaches have previously been used to study other systems of multiple detunings \cite{everettDynamicalObservationsSelfstabilizing2017,campbellDirectImagingSlow2017,vernaz-grisHighperformanceRamanMemory2018b,everettStationaryLightAtomic2019}. 

To understand the interaction of GEM and EIT in a single experiment we simulate the operation of the memory by numerical integration of the OBE using the XMDS2 package \cite{dennisXMDS2FastScalable2013} in one spatial dimension plus time (see Methods  for details). The results of these simulations are shown in Fig. \ref{fig:concept}(c-e). The original signal is stored with GEM and the TFFT of the signal is recalled with EIT. So that EIT is able to recall light from the atoms, it is crucial that the atomic spinwave is centred around $k=0$ in momentum space. In these simulations we achieve this condition artificially (see Methods for details) by launching the spinwave at negative $k$. In the experiment, as discussed later, we steer the spinwave to $k=0$ using the atomic frequency gradient.

The timings of the input and output optical signals are shown in Fig.~\ref{fig:concept}c.  We see that a pair of Gaussian pulses is recalled as a modulated Gaussian pulse and vice versa. As anticipated, this indicates a Fourier relationship between the input and output. A high optical depth of 2000 was used to reduce EIT dispersion, and a large frequency gradient was used in GEM to span the entire bandwidth of the input light. With these conditions, the recall envelope is close to a perfect Fourier transform of the input, as can be seen with the comparisons in Fig. \ref{fig:concept}(c). Some leakage of the input is evident in the data, which is a result of the large bandwidth, since the available optical depth more widely across the frequency spectrum.

By looking at the atomic coherence during storage, it is possible to see how the TFFT is realised in our system. We plot the intensity of the coherence between the F=1 and F=2 ground states, which we refer to as the spinwave.
When storing with GEM, the temporal shape of the input pulse is mapped into the momentum distribution of the spinwave \cite{hetetMultimodalPropertiesDynamics2008}, owing to the Fourier nature of the memory.  In Fig.~\ref{fig:concept}d we plot the momentum-time evolution of the spinwaves. The diagonal evolution occurs while the atomic gradient is on. Once switched off, the evolution stops and the momentum remains constant.  Taking cross-sections (insets) shows that in momentum space, the GEM spinwave envelopes have the same shape as  the input pulses.
In Fig.~\ref{fig:concept}e we plot the spinwave again, but this time in real space. The shape of the spinwave in real space is the Fourier transform of the input pulse profile \cite{hetetMultimodalPropertiesDynamics2008}. The cross sections in Fig.~\ref{fig:concept}e are therefore the Fourier transforms of the insets in Fig.~\ref{fig:concept}d. Having stored with GEM, the second stage of the TFFT is to recall with EIT. When we apply EIT conditions to this spinwave (at around 13$\mu$s), it starts to evolve and exits the memory as pulses that have a temporal envelope given by the Fourier transform of the input.

The simulations shown in the left column of Fig.~\ref{fig:concept} constitute a temporal double slit. The delay between the input pulses is transformed into the temporal frequency of the amplitude modulation of the output pulse. This is analogous to the space between a spatial double slit being mapped into the spatial frequency of the interference pattern. In both the spatial and temporal cases, interference may only occur if there is no way of knowing the path of any given photon. For a temporal double slit to be observed, there must be a way of erasing the `which-pulse' information of the photons.

The mechanism by which this occurs in our scheme is apparent in Fig. \ref{fig:concept}e. The spinwave from the second pulse is spatially superposed with the first, and the time difference between the pulses translates to a momentum difference, determining the spatial frequency of the interference. We will return to this discussion later in the context of experimental results.%

\section{Experimental demonstration of GEM-EIT}

\begin{figure}
    \centering
    \includegraphics[width=0.8\linewidth]{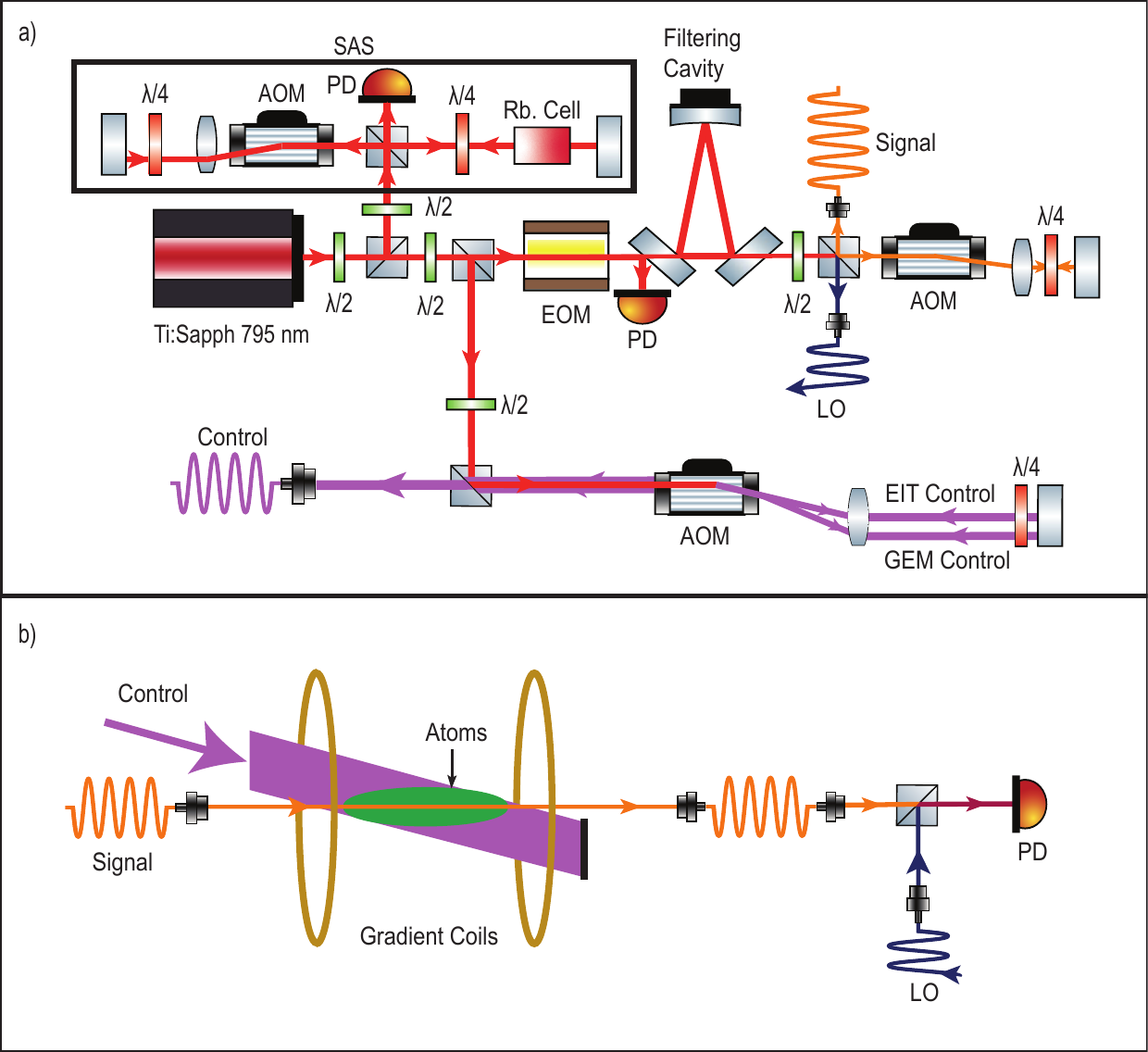}
    \caption{Schematic of the complete experimental system. a) Setup of the laser table showing the production of the two control fields, the signal and the local oscillator. AOM: acousto-optic modulator, LO: Local Oscillator, EOM: Electro-Optic Modulator, APD: Avalanche Photo Diode,  SAS: Saturated Absorption Setup, Rb. Cell: Rubidium Cell, $\lambda$/2 : half wave plate, $\lambda$/4: Quarter wave plate b) Configuration of the cold Rb ensemble and optical setup for the heterodyne detection of the transformed signal.}
    \label{fig:Setup}
\end{figure}

\begin{figure}
\centering
\includegraphics[width=1\linewidth]{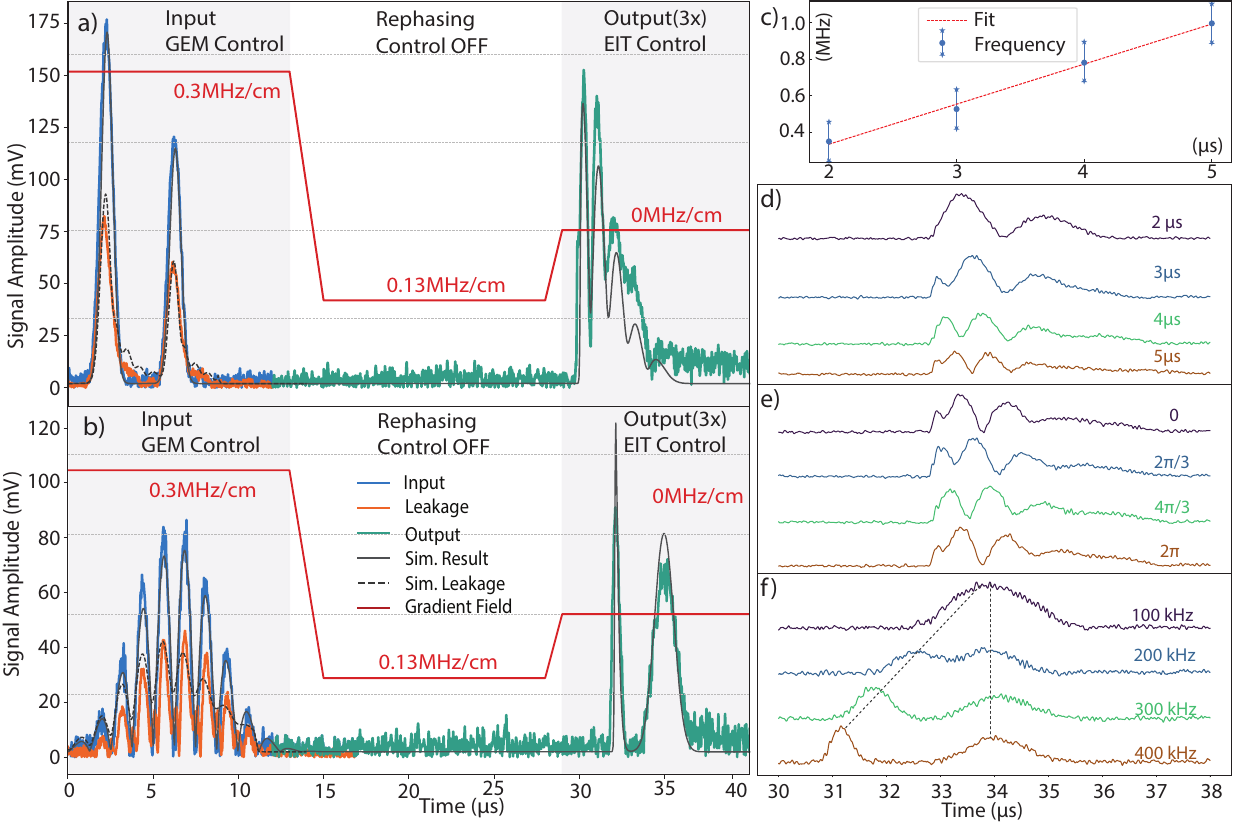}
\caption{
Input and output signals for a double-Gaussian (a) and modulated Gaussian (b). The GEM storage occurs from $0$–$13~\mu$s, rephasing ($13$–$29~\mu$s), and output ($\geq 29~\mu$s) windows; simulations are overlaid. (b) Modulated Gaussian input and output, with simulations overlaid. (c) The linear relation between frequency and pulse separation. The stars indicate the range of the chirp in the fringes. (d) Outputs for double-Gaussian inputs with varying pulse separations. (e) Outputs for double-Gaussian inputs with varying relative phases (separation fixed at $4~\mu$s). The output modulation phase follows the relative phase of the input pulses (f) Outputs for modulated Gaussian inputs at different modulation frequencies. The time separation of the output pulses increases linearly with the input modulation frequency (dotted line).}
\label{fig:Signal}
\end{figure}

The experimental realisation of this scheme is presented in Fig.~\ref{fig:Setup}. An elongated ensemble of cold thermal $^{87}$Rb atoms is prepared in a  magneto-optical trap (MOT) according to the method in Cho et al. \cite{choHighlyEfficientOptical2016}. For GEM storage, a longitudinal magnetic field gradient is applied to create the required atomic frequency gradient (see Fig.~\ref{fig:concept}a), and a signal is sent 100~MHz blue-detuned from the $\mathrm{F=1, m_F=+1 \rightarrow F'=2, m_F'=+2}$ D1 transition. A control field 100~MHz blue detuned from the $\mathrm{F=2, m_F=+1 \rightarrow F'=2, m_F'=+2}$ D1 transition is sent along with the signal, allowing 2-photon absorption of the signal. The gradient is then reversed to rephase the spinwave, then switched off. This leaves the atoms in a state where application of a control field will lead to emission of signal light. For EIT recall, a control field resonant with the $\mathrm{F=2, m_F=+1 \rightarrow F'=2, m_F'=+2}$ D1 transition (see Fig.~\ref{fig:concept}b) is used to generate the required slow light conditions.  The output signal is combined with a detuned local oscillator (Fig.~\ref{fig:Setup}b) to allow heterodyne measurement (see Methods for details). The timing of the pulses, control fields and magnetic field gradients are indicated in Fig.~\ref{fig:Signal}(a) and (b).

The experimental results of our scheme are presented in Fig.~\ref{fig:Signal}. Part (a) shows data for a double pulse input. As the memory has loss that increases with storage time, the amplitude of the input pulses was adjusted so that their recalled amplitudes were equal. The output in this case shows a single, amplitude modulated pulse. Figure~\ref{fig:Signal}b shows the data for a modulated Gaussian input. The output in this case shows a double pulse pattern.

The results are compared to numerical simulations that include the experimental parameters for optical depth, atomic cloud spatial profile and decoherence rates (see Methods for details). We note that there are numerous factors that remain unaccounted for in the model such as stray magnetic fields, inhomogeneous control field intensity, and transverse momentum of the light and spinwave.

While the results seen in Figs.~\ref{fig:Signal} (a) and (b) are in qualitative agreement with our exceptions from Fourier theory, we observe some clear differences from the ideal TFFT. In particular, we see exponential decay of output amplitude with time and  broadening of modulation for longer storage times.  These differences can be explained by the normal behaviour of GEM and EIT memory under the conditions of our experiment. Indeed, the simulations reproduce the frequency chirp and decay present in the output with reason. The major contributing factor to the broadening is the combination of  low optical depth and large time-bandwidth product. This leads to a relatively narrow EIT window, reducing the recall efficiency and significant temporal broadening of the recalled pulses. We found that by tuning the amplitude and frequency of the double-Gaussian input pulses we were able to  improve the visibility of the output pulse modulation.

The time separation of the double-Gaussian signal and the frequency of the modulated Gaussian were varied to characterise the system’s response. Figure~\ref{fig:Signal}c shows the linear relationship between the time separation of input pulses and the modulation frequency of the interference fringes, which are shown in Fig.~\ref{fig:Signal}d. These fringes have a frequency chirp imposed by the memory. The circles and stars in Fig.~~\ref{fig:Signal}c show the centre frequency and range of the chirp respectively. These values were extracted by fitting individual pulses stored by the memory (see Methods for details).  The relative phase between the input double-Gaussian pulses was also varied, to demonstrate that the phase dependence of the output modulation (Fig. \ref{fig:Signal}e). For the modulated Gaussian input, time separation of the output pulses increased linearly with modulation frequency as indicated on the diagram (Fig.~\ref{fig:Signal}f).

Previous realisations of photonic temporal double slit  experiments used spectrometers to measure the expected pattern of fringes that arise in frequency space \cite{tiroleDoubleslitTimeDiffraction2023,hongDelayedchoiceQuantumErasure2023}. To draw an analogy with spatial double slit interference it is instructive to understand where and how interference occurs between the two pulses which were initially separated in time. In a spatial double slit, the key to seeing the interference pattern is a lack of `which-slit' information.
The individual diffraction patterns from the two slits overlap in the far-field or Fourier plane, so there is no way of knowing through which slit any given photon has passed. The period of the resultant interference fringes depends on the distance between the slits.
In the case of a temporal double pulse analysed with a spectrometer, dispersion is used to broaden the individual pulses in time so much that they overlap and, with the which-pulse information now erased, interference may be observed. The modulation frequency in the interference pattern depends on the temporal separation of the pulses, analogous to the spatial double slit.

In our case, we do not require a spectrometer to observe interference. Successive pulses stored using GEM overlap spatially in the atomic ensemble. They interfere and create a spatial fringe pattern within the atomic spinwave. The spacing of the fringes is determined by the temporal separation of the pulses and the atomic frequency gradient, which adds a spatially varying phase to the spinwave over time. We normally take this intra-memory spatial interference for granted, since using GEM for both storage and recall amounts to a double Fourier transform meaning the output is just a scaled version of the input. The interference only becomes apparent when the spinwave is retrieved with EIT, and the fringe pattern appears as a function of time in the output light. In this case, the two pulses that were initially separated in time emerge together with their which-pulse information erased, so that interference fringes are now observed in the temporal profile of the output light.

As waves propagate away from a pair of spatial double slits, the diffraction pattern limits to a Fourier transform in the far-field regime. In the near-field region, however, Fresnel diffraction produces patterns that correspond to partial erasure of which-slit information. This pattern can be described using a fractional Fourier transform \cite{pellat-finetFresnelDiffractionFractionalorder1994}.  A time-frequency version of this has also been realised using an atomic memory \cite{nieweltExperimentalImplementationOptical2023}, and shown to be useful for sorting temporal modes \cite{joshiPicosecondresolutionSinglephotonTime2022}. 

A key element of TFFT methods is a `time-lens' \cite{kolnerTemporalImagingTime1989}.  In a spatial interference scenario, the far-field interference pattern can be realised at a finite distance by using a lens, which applies a spatial frequency shift to the light in proportion to the distance from the optical axis. Pure dispersion will, in principle, only realise a TFFT in the limit of infinite propagation distance.
Analogous to the spatial lens, therefore, a time-lens applies a temporal frequency shift proportional to time so that finite dispersion will be able to realise a Fourier transform. For example in \cite{joshiPicosecondresolutionSinglephotonTime2022}, the time-lens is applied by a nonlinear fibre-based dispersion, in between sections of regular dispersive fibre.

In our work, a time-lens is also present since the different arrival times when storing with GEM are converted to different frequencies when retrieving with EIT.  In contrast to other schemes, where the dispersion is applied separately to the time-lens, in our scheme the memory adds dispersion concurrently. Furthermore, additional dispersion or chirps could easily be added to the memory (as in Niewelt et al. \cite{nieweltExperimentalImplementationOptical2023}) to allow for applications in quantum sensing, by manipulating signals for enhanced sensitivity \cite{ shahFrequencySuperresolutionSpectrotemporal2021,lipkaSuperresolutionUltrafastPulses2024} and increasing quantum communication bandwidth through temporal mode sorting \cite{mowerHighdimensionalQuantumKey2013, serinoProgrammableTimefrequencyModesorting2025}.

In summary, our work demonstrates a  hybrid atomic quantum memory that implements a time-frequency Fourier transform on stored light. Our work expands on previous demonstrations of temporal double slits by including the time-lens and dispersion elements within the atomic system, in close analogy with spatial double slit experiments. We anticipate applications of this concept in quantum information systems and quantum sensing, since our memory is compatible with the storage and recall of quantum states.\\[2em]

\section{Methods
}%

\subsection{Modelling}
The memory behaviour was modelled using the eXtensible Multi-Dimensional Simulator (XMDS2) \cite{dennisXMDS2FastScalable2013}. Numerical solutions of the 3 Level Optical Bloch equations (OBEs, Eq.~\ref{gs}-\ref{dz}) \cite{gorshkovPhotonStorageType2007a} in one spatial dimension were obtained. We have used this method extensively in the past in the modelling of both Gradient Echo Memory (GEM) and Electromagnetically Induced Transparency (EIT) \cite{everettDynamicalObservationsSelfstabilizing2017,campbellDirectImagingSlow2017,vernaz-grisHighperformanceRamanMemory2018b,everettStationaryLightAtomic2019}.
The time derivatives (Eqs.~\ref{gs},\ref{ge}) were integrated with an Adaptive Runge-Kutta algorithm. Cross-propagation with order 4 Runge-Kutta was used to integrate the spatial derivative (Eq.~\ref{dz}) and generate the signal field from the left boundary. 

\begin{figure}[h]
    \centering
    \includegraphics[width=0.3\linewidth]{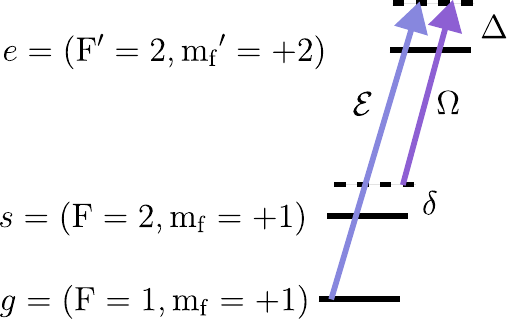}
    \caption{Level scheme for the modelling. The signal $\mathcal{E}$ and the control field $\Omega$ couple to the atomic levels shown, and the OBEs describe the evolution of $\mathcal{E}$ and the atomic coherences $\sigma_{gs}$, $\sigma_{ge}$.}
    \label{fig:levelscheme}
\end{figure}

\begin{align}
    \partial_t \hat\sigma_{ge} = ig\mathcal{E} + i\Omega \hat\sigma_{gs} - (\gamma_{ge} + i\Delta) \hat\sigma_{ge}\label{gs}\\
    \partial_t \hat\sigma_{gs} = i\Omega^* \hat\sigma_{ge} - (\gamma_{ge} + i\delta(z) - i\delta_{AC Stark}) \hat\sigma_{gs}\label{ge}\\
    \partial_z \mathcal{E} = \frac{igN}{c} \sigma_{ge}\label{dz}
\end{align}

The terms in the OBEs are: the optical coherence $\sigma_{ge}$, the spinwave coherence $\sigma_{gs}$, the decay rate for the optical coherence $\gamma_{ge}$, the signal envelope $\mathcal{E}$, the control field Rabi frequency $\Omega$, the atomic density $N$, speed of light $c$, coupling strength $g$, the 1-photon detuning $\Delta$, the 2-photon detuning (due to the gradient) $\delta(z)$, and a term compensating for AC-Stark shift $\delta_{ACStark}$.

\subsubsection{Experimental modelling}
The space and time settings were based on experimentally determined parameters. A MOT length of 30mm was divided over a lattice of 256 points. 2500 samples were taken over 100 $\mu$s evolution with ARK45, using a tolerance of $10^{-9}$.

The experimental timings, including control field and gradient switching times were taken from experimental parameters, as was single photon detuning frequency ($\Delta$). The signal pulse ($\mathcal{E}$) shape and timings for both the modulated Gaussian and double Gaussian pulses were made to closely match the experimental data.

The density distribution of the atomic cloud ($N$) was estimated using optical density measurements and transverse imaging of the atom cloud. 

The gradient magnetic field strength in the simulations was estimated using knowledge of the coil geometry, current measurements and the Raman absorption spectrum. Given that there is significant uncertainty in the estimation of these parameters (especially coil geometry), we allowed some fine-tuning to maximise agreement with experimental data. 

Optical pumping efficiency was included in the simulation by including populations and state equations for the $m_f=-1,0$ hyperfine levels. The effect of different pumping efficiencies was simulated, and a pumping efficiency of 98\% of atoms in the desired $m_f=+1$ level fit the experimental data well.

The control field Rabi frequency $\Omega$ was largely determined by the other parameters and constrained by the leakage.

\subsubsection{Idealised model}
The simulation for the idealised model operated similarly. The spatial dimension was 1 unit in length on a lattice of 800 points, with timing of 1250 samples over 25 $\mu$s using ARK89 with tolerance $10^{-5}$. 

The parameters in the theory modelling were based on an ideal ensemble of a spatially homogeneous optical depth totalling 2000, with perfect pumping into $m_f=+1$.  Some decay of the spinwave is intrinsic to the three-level model, caused mainly by the control field populating the excited state coherence.

A spatial frequency $e^{ik_0z}$ was added to the GEM control field to write the spinwave in at a momentum of $-k_0$, rather than zero. This removed the need to reverse the gradient following storage to readout with EIT at zero momentum. This makes the symmetry between the spatial envelope and momentum envelope and their evolution in the two schemes much more obvious. It was not possible to add this spatial frequency in the experiment, so the gradient was reversed instead. Momentum corresponds to frequency in EIT. A large momentum moves the signal frequency outside the EIT transparency window, causing loss. Therefore, centering the spinwave at zero momentum is important for efficient recall.

\subsection{Data Acquisition and Analysis}

We perform heterodyne detection in our setup. The local oscillator is 50 MHz red-detuned from the input signal and 50 MHz blue-detuned from the transformed signal, owing to the 100 MHz shift in control field frequency. The acquired signals were recorded using a sampling rate of 1\,GS/s and structured into two primary datasets: double Gaussian inputs and modulated-Gaussian inputs. For the double Gaussian input, data were recorded for six different pulse separations ranging from $2\,\mu\mathrm{s}$ to $5\,\mu\mathrm{s}$. For each separation, the relative phase between the two pulses was systematically varied from $0$ to $2\pi$ in steps of $\pi/3$ and both input and output references for individual pulses were recorded. In the modulated-pulse dataset, measurements were taken at 4 modulation frequencies (100--400\,kHz), with the relative phase varied from $0$ to $2\pi$ in steps of $\pi/3$ for each frequency. Each measurement cycle consisted of five input reference traces (in the absence of an atomic cloud) followed by thirty retrieval traces.

\begin{figure}
\centering
\includegraphics[width=0.8\linewidth]{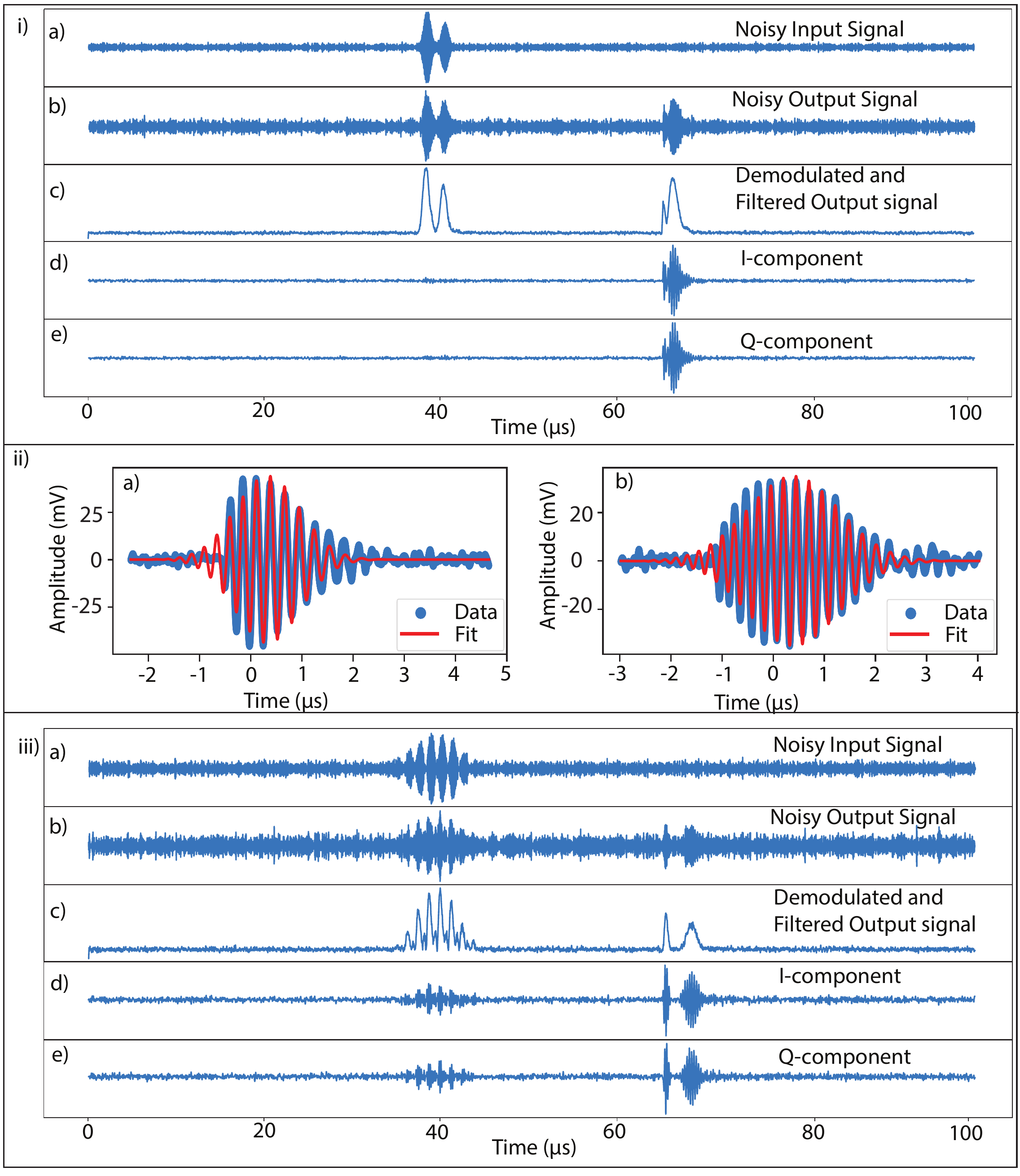}
\caption{
(i) Demodulation cycle for a double Gaussian input with a pulse separation of 2~$\mu$s. (a), (b) Single-shot traces of the unfiltered input and output signals, respectively, recorded by the oscilloscope. (c) Filtered and demodulated output corresponding to the same shot as in (b). (d), (e) In-phase (I) and quadrature (Q) components of the phase-corrected output signal, averaged over 30 shots. (ii) Processed in-phase (I) components of the outputs for the two individual references, together with their corresponding non-linear fits; the Gaussians are centred at zero to improve the accuracy and consistency of the fitting. (iii) Demodulation cycle for the modulated Gaussian input with a modulation frequency of 300~kHz, following the same structure as in (i).
}
\label{fig:DGWaveform}
\end{figure}

The signal processing workflow for both double Gaussian and modulated Gaussian inputs is summarised in Figure~\ref{fig:DGWaveform}.
To analyse the fringes resulting from the double-Gaussian input, we also fit chirped Gaussian pulses to our data, as we will describe below.

To extract the signal, we performed IQ demodulation on the heterodyne data, from which amplitude and phase were calculated using the resulting $I$ and $Q$ signals. Raw signals acquired from the oscilloscope contain not only the desired signal at the heterodyne beat frequency (typically 45–50MHz), but also broadband electronic noise, DC drift, and additional frequency components originating from the environment and measurement electronics. To isolate the signal of interest, we applied a Butterworth bandpass filter centered on the expected carrier frequency ($f_c$), with a bandwidth of ±15 MHz. This effectively preserves the main signal while suppressing low-frequency drift and high-frequency noise, thereby enhancing the fidelity of subsequent demodulation and analysis. Because jitter in each shot of a dataset introduces shot-to-shot phase variations, we corrected these by referencing all traces within each group to the first shot, ensuring accurate phase alignment across the dataset. For each experimental condition, the in-phase (I) and quadrature (Q) traces were averaged, and the resulting signal was fitted to extract the desired information.

To analyse the double Gaussian input data, we focused on determining the modulation frequency of the interference pattern observed at the output. The interference fringes in our system arise from the frequency difference between the recalled pulses. The larger the input time separation of the pulses, the larger the corresponding fringe frequency in the interference. In addition to overlapping the pulses and observing these fringes, we also stored and recalled the individual pulses in a separate series of measurements. We are then able to fit these individual pulses allowing us to find the frequency difference between the pulses that leads to the fringes. Each reference output was modelled as the real part of a complex Gaussian, with frequency, phase, and quadratic chirp (i.e. a frequency that changes linearly with time) as fitting parameters, enabling accurate extraction of the central frequency for each pulse (Fig.~\ref{fig:DGWaveform}(ii)). Since the two pulses entered the memory at different times, the difference between the extracted central frequencies of their outputs was interpreted as the modulation frequency of the double-pulse output. The range of the chirp  in the modulation frequency was calculated as 
\[
\Delta_{+} = \text{HWHM} \times \text{chirp}_1,
\]
\[
\Delta_{-} = \text{HWHM} \times \text{chirp}_2,
\]
where
\[
\text{HWHM} = 2.335 \times \frac{\sigma_\mathrm{res}}{2}, \quad
\sigma_\mathrm{res} = \sqrt{ \frac{\sigma_1^2 + \sigma_2^2}{2} }.
\]
is the Half Width at Half Maximum (HWHM), obtained from the two fitted Gaussian widths,  and 
$\text{chirp}_1$, $\text{chirp}_2$ are the extracted quadratic chirps of the first and second individual pulses, respectively. The range of frequencies in the fringe pattern ($\Delta_f$) can then be estimated as 
\[
\Delta_f= \Delta_{+} - \Delta_{-}
\]
It is this range of frequencies that we indicate in our results showing a linear relationship between pulse separation and fringe frequency.

\noindent \textbf{Acknowledgments}
This work was funded by the Australian Research Council  Center of Excellence for Quantum Computation and Communication Technology (CE170100012).

\noindent \textbf{Author contributions}\\
The modelling work was carried out by J.E. and C.T.. The experiment was conceived by B.C.B, J.E. and A.T. and carried out by A.P., J.E. and A.T.. The manuscript was drafted by all authors.

\noindent \textbf{Competing interests}\\
None.

\bibliography{main}

\begin{thebibliography}{44}%
\makeatletter
\providecommand \@ifxundefined [1]{%
 \@ifx{#1\undefined}
}%
\providecommand \@ifnum [1]{%
 \ifnum #1\expandafter \@firstoftwo
 \else \expandafter \@secondoftwo
 \fi
}%
\providecommand \@ifx [1]{%
 \ifx #1\expandafter \@firstoftwo
 \else \expandafter \@secondoftwo
 \fi
}%
\providecommand \natexlab [1]{#1}%
\providecommand \enquote  [1]{``#1''}%
\providecommand \bibnamefont  [1]{#1}%
\providecommand \bibfnamefont [1]{#1}%
\providecommand \citenamefont [1]{#1}%
\providecommand \href@noop [0]{\@secondoftwo}%
\providecommand \href [0]{\begingroup \@sanitize@url \@href}%
\providecommand \@href[1]{\@@startlink{#1}\@@href}%
\providecommand \@@href[1]{\endgroup#1\@@endlink}%
\providecommand \@sanitize@url [0]{\catcode `\\12\catcode `\$12\catcode `\&12\catcode `\#12\catcode `\^12\catcode `\_12\catcode `\%12\relax}%
\providecommand \@@startlink[1]{}%
\providecommand \@@endlink[0]{}%
\providecommand \url  [0]{\begingroup\@sanitize@url \@url }%
\providecommand \@url [1]{\endgroup\@href {#1}{\urlprefix }}%
\providecommand \urlprefix  [0]{URL }%
\providecommand \Eprint [0]{\href }%
\providecommand \doibase [0]{https://doi.org/}%
\providecommand \selectlanguage [0]{\@gobble}%
\providecommand \bibinfo  [0]{\@secondoftwo}%
\providecommand \bibfield  [0]{\@secondoftwo}%
\providecommand \translation [1]{[#1]}%
\providecommand \BibitemOpen [0]{}%
\providecommand \bibitemStop [0]{}%
\providecommand \bibitemNoStop [0]{.\EOS\space}%
\providecommand \EOS [0]{\spacefactor3000\relax}%
\providecommand \BibitemShut  [1]{\csname bibitem#1\endcsname}%
\let\auto@bib@innerbib\@empty
\bibitem [{\citenamefont {{van Howe}}\ and\ \citenamefont {Xu}(2006)}]{vanhoweUltrafastOpticalSignal2006}%
  \BibitemOpen
  \bibfield  {author} {\bibinfo {author} {\bibfnamefont {J.}~\bibnamefont {{van Howe}}}\ and\ \bibinfo {author} {\bibfnamefont {C.}~\bibnamefont {Xu}},\ }\bibfield  {title} {\bibinfo {title} {Ultrafast optical signal processing based upon space-time dualities},\ }\href {https://doi.org/10.1109/JLT.2006.875229} {\bibfield  {journal} {\bibinfo  {journal} {Journal of Lightwave Technology}\ }\textbf {\bibinfo {volume} {24}},\ \bibinfo {pages} {2649} (\bibinfo {year} {2006})}\BibitemShut {NoStop}%
\bibitem [{\citenamefont {Tsang}\ and\ \citenamefont {Psaltis}(2006)}]{tsangPropagationTemporalEntanglement2006}%
  \BibitemOpen
  \bibfield  {author} {\bibinfo {author} {\bibfnamefont {M.}~\bibnamefont {Tsang}}\ and\ \bibinfo {author} {\bibfnamefont {D.}~\bibnamefont {Psaltis}},\ }\bibfield  {title} {\bibinfo {title} {Propagation of temporal entanglement},\ }\href {https://doi.org/10.1103/PhysRevA.73.013822} {\bibfield  {journal} {\bibinfo  {journal} {Physical Review A}\ }\textbf {\bibinfo {volume} {73}},\ \bibinfo {pages} {013822} (\bibinfo {year} {2006})}\BibitemShut {NoStop}%
\bibitem [{\citenamefont {Donohue}\ \emph {et~al.}(2016)\citenamefont {Donohue}, \citenamefont {Mastrovich},\ and\ \citenamefont {Resch}}]{donohueSpectrallyEngineeringPhotonic2016}%
  \BibitemOpen
  \bibfield  {author} {\bibinfo {author} {\bibfnamefont {J.~M.}\ \bibnamefont {Donohue}}, \bibinfo {author} {\bibfnamefont {M.}~\bibnamefont {Mastrovich}},\ and\ \bibinfo {author} {\bibfnamefont {K.~J.}\ \bibnamefont {Resch}},\ }\bibfield  {title} {\bibinfo {title} {Spectrally {{Engineering Photonic Entanglement}} with a {{Time Lens}}},\ }\href {https://doi.org/10.1103/PhysRevLett.117.243602} {\bibfield  {journal} {\bibinfo  {journal} {Physical Review Letters}\ }\textbf {\bibinfo {volume} {117}},\ \bibinfo {pages} {243602} (\bibinfo {year} {2016})}\BibitemShut {NoStop}%
\bibitem [{\citenamefont {Matsuda}(2016)}]{matsudaDeterministicReshapingSinglephoton2016}%
  \BibitemOpen
  \bibfield  {author} {\bibinfo {author} {\bibfnamefont {N.}~\bibnamefont {Matsuda}},\ }\bibfield  {title} {\bibinfo {title} {Deterministic reshaping of single-photon spectra using cross-phase modulation},\ }\href {https://doi.org/10.1126/sciadv.1501223} {\bibfield  {journal} {\bibinfo  {journal} {Science Advances}\ }\textbf {\bibinfo {volume} {2}},\ \bibinfo {pages} {e1501223} (\bibinfo {year} {2016})}\BibitemShut {NoStop}%
\bibitem [{\citenamefont {Karpi{\'n}ski}\ \emph {et~al.}(2017)\citenamefont {Karpi{\'n}ski}, \citenamefont {Jachura}, \citenamefont {Wright},\ and\ \citenamefont {Smith}}]{karpinskiBandwidthManipulationQuantum2017}%
  \BibitemOpen
  \bibfield  {author} {\bibinfo {author} {\bibfnamefont {M.}~\bibnamefont {Karpi{\'n}ski}}, \bibinfo {author} {\bibfnamefont {M.}~\bibnamefont {Jachura}}, \bibinfo {author} {\bibfnamefont {L.~J.}\ \bibnamefont {Wright}},\ and\ \bibinfo {author} {\bibfnamefont {B.~J.}\ \bibnamefont {Smith}},\ }\bibfield  {title} {\bibinfo {title} {Bandwidth manipulation of quantum light by an electro-optic time lens},\ }\href {https://doi.org/10.1038/nphoton.2016.228} {\bibfield  {journal} {\bibinfo  {journal} {Nature Photonics}\ }\textbf {\bibinfo {volume} {11}},\ \bibinfo {pages} {53} (\bibinfo {year} {2017})}\BibitemShut {NoStop}%
\bibitem [{\citenamefont {Lu}\ \emph {et~al.}(2018)\citenamefont {Lu}, \citenamefont {Lukens}, \citenamefont {Peters}, \citenamefont {Williams}, \citenamefont {Weiner},\ and\ \citenamefont {Lougovski}}]{luQuantumInterferenceCorrelation2018}%
  \BibitemOpen
  \bibfield  {author} {\bibinfo {author} {\bibfnamefont {H.-H.}\ \bibnamefont {Lu}}, \bibinfo {author} {\bibfnamefont {J.~M.}\ \bibnamefont {Lukens}}, \bibinfo {author} {\bibfnamefont {N.~A.}\ \bibnamefont {Peters}}, \bibinfo {author} {\bibfnamefont {B.~P.}\ \bibnamefont {Williams}}, \bibinfo {author} {\bibfnamefont {A.~M.}\ \bibnamefont {Weiner}},\ and\ \bibinfo {author} {\bibfnamefont {P.}~\bibnamefont {Lougovski}},\ }\bibfield  {title} {\bibinfo {title} {Quantum interference and correlation control of frequency-bin qubits},\ }\href {https://doi.org/10.1364/OPTICA.5.001455} {\bibfield  {journal} {\bibinfo  {journal} {Optica}\ }\textbf {\bibinfo {volume} {5}},\ \bibinfo {pages} {1455} (\bibinfo {year} {2018})}\BibitemShut {NoStop}%
\bibitem [{\citenamefont {Francesconi}\ \emph {et~al.}(2020)\citenamefont {Francesconi}, \citenamefont {Baboux}, \citenamefont {Raymond}, \citenamefont {Fabre}, \citenamefont {Boucher}, \citenamefont {Lema{\^i}tre}, \citenamefont {Milman}, \citenamefont {Amanti},\ and\ \citenamefont {Ducci}}]{francesconiEngineeringTwophotonWavefunction2020}%
  \BibitemOpen
  \bibfield  {author} {\bibinfo {author} {\bibfnamefont {S.}~\bibnamefont {Francesconi}}, \bibinfo {author} {\bibfnamefont {F.}~\bibnamefont {Baboux}}, \bibinfo {author} {\bibfnamefont {A.}~\bibnamefont {Raymond}}, \bibinfo {author} {\bibfnamefont {N.}~\bibnamefont {Fabre}}, \bibinfo {author} {\bibfnamefont {G.}~\bibnamefont {Boucher}}, \bibinfo {author} {\bibfnamefont {A.}~\bibnamefont {Lema{\^i}tre}}, \bibinfo {author} {\bibfnamefont {P.}~\bibnamefont {Milman}}, \bibinfo {author} {\bibfnamefont {M.~I.}\ \bibnamefont {Amanti}},\ and\ \bibinfo {author} {\bibfnamefont {S.}~\bibnamefont {Ducci}},\ }\bibfield  {title} {\bibinfo {title} {Engineering two-photon wavefunction and exchange statistics in a semiconductor chip},\ }\href {https://doi.org/10.1364/OPTICA.379477} {\bibfield  {journal} {\bibinfo  {journal} {Optica}\ }\textbf {\bibinfo {volume} {7}},\ \bibinfo {pages} {316} (\bibinfo {year} {2020})}\BibitemShut {NoStop}%
\bibitem [{\citenamefont {So{\'s}nicki}\ \emph {et~al.}(2020)\citenamefont {So{\'s}nicki}, \citenamefont {Miko{\l}ajczyk}, \citenamefont {Golestani},\ and\ \citenamefont {Karpi{\'n}ski}}]{sosnickiAperiodicElectroopticTime2020}%
  \BibitemOpen
  \bibfield  {author} {\bibinfo {author} {\bibfnamefont {F.}~\bibnamefont {So{\'s}nicki}}, \bibinfo {author} {\bibfnamefont {M.}~\bibnamefont {Miko{\l}ajczyk}}, \bibinfo {author} {\bibfnamefont {A.}~\bibnamefont {Golestani}},\ and\ \bibinfo {author} {\bibfnamefont {M.}~\bibnamefont {Karpi{\'n}ski}},\ }\bibfield  {title} {\bibinfo {title} {Aperiodic electro-optic time lens for spectral manipulation of single-photon pulses},\ }\href {https://doi.org/10.1063/5.0011077} {\bibfield  {journal} {\bibinfo  {journal} {Applied Physics Letters}\ }\textbf {\bibinfo {volume} {116}},\ \bibinfo {pages} {234003} (\bibinfo {year} {2020})}\BibitemShut {NoStop}%
\bibitem [{\citenamefont {Joshi}\ \emph {et~al.}(2022)\citenamefont {Joshi}, \citenamefont {Sparkes}, \citenamefont {Farsi}, \citenamefont {Gerrits}, \citenamefont {Verma}, \citenamefont {Ramelow}, \citenamefont {Nam},\ and\ \citenamefont {Gaeta}}]{joshiPicosecondresolutionSinglephotonTime2022}%
  \BibitemOpen
  \bibfield  {author} {\bibinfo {author} {\bibfnamefont {C.}~\bibnamefont {Joshi}}, \bibinfo {author} {\bibfnamefont {B.~M.}\ \bibnamefont {Sparkes}}, \bibinfo {author} {\bibfnamefont {A.}~\bibnamefont {Farsi}}, \bibinfo {author} {\bibfnamefont {T.}~\bibnamefont {Gerrits}}, \bibinfo {author} {\bibfnamefont {V.}~\bibnamefont {Verma}}, \bibinfo {author} {\bibfnamefont {S.}~\bibnamefont {Ramelow}}, \bibinfo {author} {\bibfnamefont {S.~W.}\ \bibnamefont {Nam}},\ and\ \bibinfo {author} {\bibfnamefont {A.~L.}\ \bibnamefont {Gaeta}},\ }\bibfield  {title} {\bibinfo {title} {Picosecond-resolution single-photon time lens for temporal mode quantum processing},\ }\href {https://doi.org/10.1364/OPTICA.439827} {\bibfield  {journal} {\bibinfo  {journal} {Optica}\ }\textbf {\bibinfo {volume} {9}},\ \bibinfo {pages} {364} (\bibinfo {year} {2022})}\BibitemShut {NoStop}%
\bibitem [{\citenamefont {Lipka}\ and\ \citenamefont {Parniak}(2024)}]{lipkaSuperresolutionUltrafastPulses2024}%
  \BibitemOpen
  \bibfield  {author} {\bibinfo {author} {\bibfnamefont {M.}~\bibnamefont {Lipka}}\ and\ \bibinfo {author} {\bibfnamefont {M.}~\bibnamefont {Parniak}},\ }\bibfield  {title} {\bibinfo {title} {Super-resolution of ultrafast pulses via spectral inversion},\ }\href {https://doi.org/10.1364/OPTICA.522555} {\bibfield  {journal} {\bibinfo  {journal} {Optica}\ }\textbf {\bibinfo {volume} {11}},\ \bibinfo {pages} {1226} (\bibinfo {year} {2024})}\BibitemShut {NoStop}%
\bibitem [{\citenamefont {Mazelanik}\ \emph {et~al.}(2020)\citenamefont {Mazelanik}, \citenamefont {Leszczy{\'n}ski}, \citenamefont {Lipka}, \citenamefont {Parniak},\ and\ \citenamefont {Wasilewski}}]{mazelanikTemporalImagingUltranarrowband2020a}%
  \BibitemOpen
  \bibfield  {author} {\bibinfo {author} {\bibfnamefont {M.}~\bibnamefont {Mazelanik}}, \bibinfo {author} {\bibfnamefont {A.}~\bibnamefont {Leszczy{\'n}ski}}, \bibinfo {author} {\bibfnamefont {M.}~\bibnamefont {Lipka}}, \bibinfo {author} {\bibfnamefont {M.}~\bibnamefont {Parniak}},\ and\ \bibinfo {author} {\bibfnamefont {W.}~\bibnamefont {Wasilewski}},\ }\bibfield  {title} {\bibinfo {title} {Temporal imaging for ultra-narrowband few-photon states of light},\ }\href {https://doi.org/10.1364/OPTICA.382891} {\bibfield  {journal} {\bibinfo  {journal} {Optica}\ }\textbf {\bibinfo {volume} {7}},\ \bibinfo {pages} {203} (\bibinfo {year} {2020})}\BibitemShut {NoStop}%
\bibitem [{\citenamefont {Mazelanik}\ \emph {et~al.}(2022)\citenamefont {Mazelanik}, \citenamefont {Leszczy{\'n}ski},\ and\ \citenamefont {Parniak}}]{mazelanikOpticaldomainSpectralSuperresolution2022a}%
  \BibitemOpen
  \bibfield  {author} {\bibinfo {author} {\bibfnamefont {M.}~\bibnamefont {Mazelanik}}, \bibinfo {author} {\bibfnamefont {A.}~\bibnamefont {Leszczy{\'n}ski}},\ and\ \bibinfo {author} {\bibfnamefont {M.}~\bibnamefont {Parniak}},\ }\bibfield  {title} {\bibinfo {title} {Optical-domain spectral super-resolution via a quantum-memory-based time-frequency processor},\ }\href {https://doi.org/10.1038/s41467-022-28066-5} {\bibfield  {journal} {\bibinfo  {journal} {Nature Communications}\ }\textbf {\bibinfo {volume} {13}},\ \bibinfo {pages} {691} (\bibinfo {year} {2022})}\BibitemShut {NoStop}%
\bibitem [{\citenamefont {Niewelt}\ \emph {et~al.}(2023)\citenamefont {Niewelt}, \citenamefont {Jastrz{\k e}bski}, \citenamefont {Kurzyna}, \citenamefont {Nowosielski}, \citenamefont {Wasilewski}, \citenamefont {Mazelanik},\ and\ \citenamefont {Parniak}}]{nieweltExperimentalImplementationOptical2023}%
  \BibitemOpen
  \bibfield  {author} {\bibinfo {author} {\bibfnamefont {B.}~\bibnamefont {Niewelt}}, \bibinfo {author} {\bibfnamefont {M.}~\bibnamefont {Jastrz{\k e}bski}}, \bibinfo {author} {\bibfnamefont {S.}~\bibnamefont {Kurzyna}}, \bibinfo {author} {\bibfnamefont {J.}~\bibnamefont {Nowosielski}}, \bibinfo {author} {\bibfnamefont {W.}~\bibnamefont {Wasilewski}}, \bibinfo {author} {\bibfnamefont {M.}~\bibnamefont {Mazelanik}},\ and\ \bibinfo {author} {\bibfnamefont {M.}~\bibnamefont {Parniak}},\ }\bibfield  {title} {\bibinfo {title} {Experimental {{Implementation}} of the {{Optical Fractional Fourier Transform}} in the {{Time-Frequency Domain}}},\ }\href {https://doi.org/10.1103/PhysRevLett.130.240801} {\bibfield  {journal} {\bibinfo  {journal} {Physical Review Letters}\ }\textbf {\bibinfo {volume} {130}},\ \bibinfo {pages} {240801} (\bibinfo {year} {2023})}\BibitemShut {NoStop}%
\bibitem [{\citenamefont {Kolner}(1994)}]{kolnerSpacetimeDualityTheory1994}%
  \BibitemOpen
  \bibfield  {author} {\bibinfo {author} {\bibfnamefont {B.}~\bibnamefont {Kolner}},\ }\bibfield  {title} {\bibinfo {title} {Space-time duality and the theory of temporal imaging},\ }\href {https://doi.org/10.1109/3.301659} {\bibfield  {journal} {\bibinfo  {journal} {IEEE Journal of Quantum Electronics}\ }\textbf {\bibinfo {volume} {30}},\ \bibinfo {pages} {1951} (\bibinfo {year} {1994})}\BibitemShut {NoStop}%
\bibitem [{\citenamefont {Karpi{\'n}ski}\ \emph {et~al.}(2021)\citenamefont {Karpi{\'n}ski}, \citenamefont {Davis}, \citenamefont {So{\'s}nicki}, \citenamefont {Thiel},\ and\ \citenamefont {Smith}}]{karpinskiControlMeasurementQuantum2021}%
  \BibitemOpen
  \bibfield  {author} {\bibinfo {author} {\bibfnamefont {M.}~\bibnamefont {Karpi{\'n}ski}}, \bibinfo {author} {\bibfnamefont {A.~O.~C.}\ \bibnamefont {Davis}}, \bibinfo {author} {\bibfnamefont {F.}~\bibnamefont {So{\'s}nicki}}, \bibinfo {author} {\bibfnamefont {V.}~\bibnamefont {Thiel}},\ and\ \bibinfo {author} {\bibfnamefont {B.~J.}\ \bibnamefont {Smith}},\ }\bibfield  {title} {\bibinfo {title} {Control and {{Measurement}} of {{Quantum Light Pulses}} for {{Quantum Information Science}} and {{Technology}}},\ }\href {https://doi.org/10.1002/qute.202000150} {\bibfield  {journal} {\bibinfo  {journal} {Advanced Quantum Technologies}\ }\textbf {\bibinfo {volume} {4}},\ \bibinfo {pages} {2000150} (\bibinfo {year} {2021})}\BibitemShut {NoStop}%
\bibitem [{\citenamefont {Lei}\ \emph {et~al.}(2023)\citenamefont {Lei}, \citenamefont {Asadi}, \citenamefont {Zhong}, \citenamefont {Kuzmich}, \citenamefont {Simon},\ and\ \citenamefont {Hosseini}}]{Lei:23}%
  \BibitemOpen
  \bibfield  {author} {\bibinfo {author} {\bibfnamefont {Y.}~\bibnamefont {Lei}}, \bibinfo {author} {\bibfnamefont {F.~K.}\ \bibnamefont {Asadi}}, \bibinfo {author} {\bibfnamefont {T.}~\bibnamefont {Zhong}}, \bibinfo {author} {\bibfnamefont {A.}~\bibnamefont {Kuzmich}}, \bibinfo {author} {\bibfnamefont {C.}~\bibnamefont {Simon}},\ and\ \bibinfo {author} {\bibfnamefont {M.}~\bibnamefont {Hosseini}},\ }\bibfield  {title} {\bibinfo {title} {Quantum optical memory for entanglement distribution},\ }\href {https://doi.org/10.1364/OPTICA.493732} {\bibfield  {journal} {\bibinfo  {journal} {Optica}\ }\textbf {\bibinfo {volume} {10}},\ \bibinfo {pages} {1511} (\bibinfo {year} {2023})}\BibitemShut {NoStop}%
\bibitem [{\citenamefont {Azuma}\ \emph {et~al.}(2023)\citenamefont {Azuma}, \citenamefont {Economou}, \citenamefont {Elkouss}, \citenamefont {Hilaire}, \citenamefont {Jiang}, \citenamefont {Lo},\ and\ \citenamefont {Tzitrin}}]{RevModPhys.95.045006}%
  \BibitemOpen
  \bibfield  {author} {\bibinfo {author} {\bibfnamefont {K.}~\bibnamefont {Azuma}}, \bibinfo {author} {\bibfnamefont {S.~E.}\ \bibnamefont {Economou}}, \bibinfo {author} {\bibfnamefont {D.}~\bibnamefont {Elkouss}}, \bibinfo {author} {\bibfnamefont {P.}~\bibnamefont {Hilaire}}, \bibinfo {author} {\bibfnamefont {L.}~\bibnamefont {Jiang}}, \bibinfo {author} {\bibfnamefont {H.-K.}\ \bibnamefont {Lo}},\ and\ \bibinfo {author} {\bibfnamefont {I.}~\bibnamefont {Tzitrin}},\ }\bibfield  {title} {\bibinfo {title} {Quantum repeaters: From quantum networks to the quantum internet},\ }\href {https://doi.org/10.1103/RevModPhys.95.045006} {\bibfield  {journal} {\bibinfo  {journal} {Rev. Mod. Phys.}\ }\textbf {\bibinfo {volume} {95}},\ \bibinfo {pages} {045006} (\bibinfo {year} {2023})}\BibitemShut {NoStop}%
\bibitem [{\citenamefont {Barral}\ \emph {et~al.}(2025)\citenamefont {Barral}, \citenamefont {Cardama}, \citenamefont {Díaz-Camacho}, \citenamefont {Faílde}, \citenamefont {Llovo}, \citenamefont {Mussa-Juane}, \citenamefont {Vázquez-Pérez}, \citenamefont {Villasuso}, \citenamefont {Piñeiro}, \citenamefont {Costas}, \citenamefont {Pichel}, \citenamefont {Pena},\ and\ \citenamefont {Gómez}}]{BARRAL2025100747}%
  \BibitemOpen
  \bibfield  {author} {\bibinfo {author} {\bibfnamefont {D.}~\bibnamefont {Barral}}, \bibinfo {author} {\bibfnamefont {F.~J.}\ \bibnamefont {Cardama}}, \bibinfo {author} {\bibfnamefont {G.}~\bibnamefont {Díaz-Camacho}}, \bibinfo {author} {\bibfnamefont {D.}~\bibnamefont {Faílde}}, \bibinfo {author} {\bibfnamefont {I.~F.}\ \bibnamefont {Llovo}}, \bibinfo {author} {\bibfnamefont {M.}~\bibnamefont {Mussa-Juane}}, \bibinfo {author} {\bibfnamefont {J.}~\bibnamefont {Vázquez-Pérez}}, \bibinfo {author} {\bibfnamefont {J.}~\bibnamefont {Villasuso}}, \bibinfo {author} {\bibfnamefont {C.}~\bibnamefont {Piñeiro}}, \bibinfo {author} {\bibfnamefont {N.}~\bibnamefont {Costas}}, \bibinfo {author} {\bibfnamefont {J.~C.}\ \bibnamefont {Pichel}}, \bibinfo {author} {\bibfnamefont {T.~F.}\ \bibnamefont {Pena}},\ and\ \bibinfo {author} {\bibfnamefont {A.}~\bibnamefont {Gómez}},\ }\bibfield  {title} {\bibinfo {title} {Review of distributed quantum computing: From single qpu to high performance quantum computing},\ }\href
  {https://doi.org/https://doi.org/10.1016/j.cosrev.2025.100747} {\bibfield  {journal} {\bibinfo  {journal} {Computer Science Review}\ }\textbf {\bibinfo {volume} {57}},\ \bibinfo {pages} {100747} (\bibinfo {year} {2025})}\BibitemShut {NoStop}%
\bibitem [{\citenamefont {Lvovsky}\ \emph {et~al.}(2009)\citenamefont {Lvovsky}, \citenamefont {Sanders},\ and\ \citenamefont {Tittel}}]{lvovskyOpticalQuantumMemory2009a}%
  \BibitemOpen
  \bibfield  {author} {\bibinfo {author} {\bibfnamefont {A.~I.}\ \bibnamefont {Lvovsky}}, \bibinfo {author} {\bibfnamefont {B.~C.}\ \bibnamefont {Sanders}},\ and\ \bibinfo {author} {\bibfnamefont {W.}~\bibnamefont {Tittel}},\ }\bibfield  {title} {\bibinfo {title} {Optical quantum memory},\ }\href {https://doi.org/10.1038/nphoton.2009.231} {\bibfield  {journal} {\bibinfo  {journal} {Nature Photonics}\ }\textbf {\bibinfo {volume} {3}},\ \bibinfo {pages} {706} (\bibinfo {year} {2009})}\BibitemShut {NoStop}%
\bibitem [{\citenamefont {Harris}(1997)}]{harrisElectromagneticallyInducedTransparency1997}%
  \BibitemOpen
  \bibfield  {author} {\bibinfo {author} {\bibfnamefont {S.~E.}\ \bibnamefont {Harris}},\ }\bibfield  {title} {\bibinfo {title} {Electromagnetically {{Induced Transparency}}},\ }\href {https://doi.org/10.1063/1.881806} {\bibfield  {journal} {\bibinfo  {journal} {Physics Today}\ }\textbf {\bibinfo {volume} {50}},\ \bibinfo {pages} {36} (\bibinfo {year} {1997})}\BibitemShut {NoStop}%
\bibitem [{\citenamefont {Hsiao}\ \emph {et~al.}(2018)\citenamefont {Hsiao}, \citenamefont {Tsai}, \citenamefont {Chen}, \citenamefont {Lin}, \citenamefont {Hung}, \citenamefont {Lee}, \citenamefont {Chen}, \citenamefont {Chen}, \citenamefont {Yu},\ and\ \citenamefont {Chen}}]{hsiaoHighlyEfficientCoherent2018a}%
  \BibitemOpen
  \bibfield  {author} {\bibinfo {author} {\bibfnamefont {Y.-F.}\ \bibnamefont {Hsiao}}, \bibinfo {author} {\bibfnamefont {P.-J.}\ \bibnamefont {Tsai}}, \bibinfo {author} {\bibfnamefont {H.-S.}\ \bibnamefont {Chen}}, \bibinfo {author} {\bibfnamefont {S.-X.}\ \bibnamefont {Lin}}, \bibinfo {author} {\bibfnamefont {C.-C.}\ \bibnamefont {Hung}}, \bibinfo {author} {\bibfnamefont {C.-H.}\ \bibnamefont {Lee}}, \bibinfo {author} {\bibfnamefont {Y.-H.}\ \bibnamefont {Chen}}, \bibinfo {author} {\bibfnamefont {Y.-F.}\ \bibnamefont {Chen}}, \bibinfo {author} {\bibfnamefont {I.~A.}\ \bibnamefont {Yu}},\ and\ \bibinfo {author} {\bibfnamefont {Y.-C.}\ \bibnamefont {Chen}},\ }\bibfield  {title} {\bibinfo {title} {Highly {{Efficient Coherent Optical Memory Based}} on {{Electromagnetically Induced Transparency}}},\ }\href {https://doi.org/10.1103/PhysRevLett.120.183602} {\bibfield  {journal} {\bibinfo  {journal} {Physical Review Letters}\ }\textbf {\bibinfo {volume} {120}},\ \bibinfo {pages} {183602} (\bibinfo {year}
  {2018})}\BibitemShut {NoStop}%
\bibitem [{\citenamefont {{Vernaz-Gris}}\ \emph {et~al.}(2018{\natexlab{a}})\citenamefont {{Vernaz-Gris}}, \citenamefont {Huang}, \citenamefont {Cao}, \citenamefont {Sheremet},\ and\ \citenamefont {Laurat}}]{vernaz-grisHighlyefficientQuantumMemory2018}%
  \BibitemOpen
  \bibfield  {author} {\bibinfo {author} {\bibfnamefont {P.}~\bibnamefont {{Vernaz-Gris}}}, \bibinfo {author} {\bibfnamefont {K.}~\bibnamefont {Huang}}, \bibinfo {author} {\bibfnamefont {M.}~\bibnamefont {Cao}}, \bibinfo {author} {\bibfnamefont {A.~S.}\ \bibnamefont {Sheremet}},\ and\ \bibinfo {author} {\bibfnamefont {J.}~\bibnamefont {Laurat}},\ }\bibfield  {title} {\bibinfo {title} {Highly-efficient quantum memory for polarization qubits in a spatially-multiplexed cold atomic ensemble},\ }\bibfield  {journal} {\bibinfo  {journal} {Nature Communications}\ }\textbf {\bibinfo {volume} {9}},\ \href {https://doi.org/10.1038/s41467-017-02775-8} {10.1038/s41467-017-02775-8} (\bibinfo {year} {2018}{\natexlab{a}})\BibitemShut {NoStop}%
\bibitem [{\citenamefont {Moiseev}\ and\ \citenamefont {Arslanov}(2008)}]{moiseevEfficiencyFidelityPhotonecho2008}%
  \BibitemOpen
  \bibfield  {author} {\bibinfo {author} {\bibfnamefont {S.~A.}\ \bibnamefont {Moiseev}}\ and\ \bibinfo {author} {\bibfnamefont {N.~M.}\ \bibnamefont {Arslanov}},\ }\bibfield  {title} {\bibinfo {title} {Efficiency and fidelity of photon-echo quantum memory in an atomic system with longitudinal inhomogeneous broadening},\ }\bibfield  {journal} {\bibinfo  {journal} {Physical Review A}\ }\textbf {\bibinfo {volume} {78}},\ \href {https://doi.org/10.1103/PhysRevA.78.023803} {10.1103/PhysRevA.78.023803} (\bibinfo {year} {2008})\BibitemShut {NoStop}%
\bibitem [{\citenamefont {H{\'e}tet}\ \emph {et~al.}(2008)\citenamefont {H{\'e}tet}, \citenamefont {Longdell}, \citenamefont {Sellars}, \citenamefont {Lam},\ and\ \citenamefont {Buchler}}]{hetetMultimodalPropertiesDynamics2008}%
  \BibitemOpen
  \bibfield  {author} {\bibinfo {author} {\bibfnamefont {G.}~\bibnamefont {H{\'e}tet}}, \bibinfo {author} {\bibfnamefont {J.~J.}\ \bibnamefont {Longdell}}, \bibinfo {author} {\bibfnamefont {M.~J.}\ \bibnamefont {Sellars}}, \bibinfo {author} {\bibfnamefont {P.~K.}\ \bibnamefont {Lam}},\ and\ \bibinfo {author} {\bibfnamefont {B.~C.}\ \bibnamefont {Buchler}},\ }\bibfield  {title} {\bibinfo {title} {Multimodal {{Properties}} and {{Dynamics}} of {{Gradient Echo Quantum Memory}}},\ }\bibfield  {journal} {\bibinfo  {journal} {Physical Review Letters}\ }\textbf {\bibinfo {volume} {101}},\ \href {https://doi.org/10.1103/PhysRevLett.101.203601} {10.1103/PhysRevLett.101.203601} (\bibinfo {year} {2008})\BibitemShut {NoStop}%
\bibitem [{\citenamefont {Campbell}\ \emph {et~al.}(2016)\citenamefont {Campbell}, \citenamefont {Ferguson}, \citenamefont {Sellars}, \citenamefont {Buchler},\ and\ \citenamefont {Lam}}]{campbellEchoBasedQuantumMemory2016}%
  \BibitemOpen
  \bibfield  {author} {\bibinfo {author} {\bibfnamefont {G.~T.}\ \bibnamefont {Campbell}}, \bibinfo {author} {\bibfnamefont {K.~R.}\ \bibnamefont {Ferguson}}, \bibinfo {author} {\bibfnamefont {M.~J.}\ \bibnamefont {Sellars}}, \bibinfo {author} {\bibfnamefont {B.~C.}\ \bibnamefont {Buchler}},\ and\ \bibinfo {author} {\bibfnamefont {P.~K.}\ \bibnamefont {Lam}},\ }\bibfield  {title} {\bibinfo {title} {Echo-{{Based Quantum Memory}}},\ }in\ \href {https://doi.org/10.1002/9783527805785.ch32} {\emph {\bibinfo {booktitle} {Quantum {{Information}}}}}\ (\bibinfo  {publisher} {John Wiley \& Sons, Ltd},\ \bibinfo {year} {2016})\ Chap.~\bibinfo {chapter} {32}, pp.\ \bibinfo {pages} {723--740}\BibitemShut {NoStop}%
\bibitem [{\citenamefont {Cho}\ \emph {et~al.}(2016)\citenamefont {Cho}, \citenamefont {Campbell}, \citenamefont {Everett}, \citenamefont {Bernu}, \citenamefont {Higginbottom}, \citenamefont {Cao}, \citenamefont {Geng}, \citenamefont {Robins}, \citenamefont {Lam},\ and\ \citenamefont {Buchler}}]{choHighlyEfficientOptical2016}%
  \BibitemOpen
  \bibfield  {author} {\bibinfo {author} {\bibfnamefont {Y.-W.}\ \bibnamefont {Cho}}, \bibinfo {author} {\bibfnamefont {G.~T.}\ \bibnamefont {Campbell}}, \bibinfo {author} {\bibfnamefont {J.~L.}\ \bibnamefont {Everett}}, \bibinfo {author} {\bibfnamefont {J.}~\bibnamefont {Bernu}}, \bibinfo {author} {\bibfnamefont {D.~B.}\ \bibnamefont {Higginbottom}}, \bibinfo {author} {\bibfnamefont {M.~T.}\ \bibnamefont {Cao}}, \bibinfo {author} {\bibfnamefont {J.}~\bibnamefont {Geng}}, \bibinfo {author} {\bibfnamefont {N.~P.}\ \bibnamefont {Robins}}, \bibinfo {author} {\bibfnamefont {P.~K.}\ \bibnamefont {Lam}},\ and\ \bibinfo {author} {\bibfnamefont {B.~C.}\ \bibnamefont {Buchler}},\ }\bibfield  {title} {\bibinfo {title} {Highly efficient optical quantum memory with long coherence time in cold atoms},\ }\href {https://doi.org/10.1364/OPTICA.3.000100} {\bibfield  {journal} {\bibinfo  {journal} {Optica}\ }\textbf {\bibinfo {volume} {3}},\ \bibinfo {pages} {100} (\bibinfo {year} {2016})}\BibitemShut {NoStop}%
\bibitem [{\citenamefont {Leung}\ \emph {et~al.}(2024)\citenamefont {Leung}, \citenamefont {Lau}, \citenamefont {Tranter}, \citenamefont {Paul}, \citenamefont {Rambach}, \citenamefont {Buchler}, \citenamefont {Lam}, \citenamefont {White},\ and\ \citenamefont {Weinhold}}]{leungHighlyEfficientStorage2024}%
  \BibitemOpen
  \bibfield  {author} {\bibinfo {author} {\bibfnamefont {A.~C.}\ \bibnamefont {Leung}}, \bibinfo {author} {\bibfnamefont {W.~Y.~S.}\ \bibnamefont {Lau}}, \bibinfo {author} {\bibfnamefont {A.~D.}\ \bibnamefont {Tranter}}, \bibinfo {author} {\bibfnamefont {K.~V.}\ \bibnamefont {Paul}}, \bibinfo {author} {\bibfnamefont {M.}~\bibnamefont {Rambach}}, \bibinfo {author} {\bibfnamefont {B.~C.}\ \bibnamefont {Buchler}}, \bibinfo {author} {\bibfnamefont {P.~K.}\ \bibnamefont {Lam}}, \bibinfo {author} {\bibfnamefont {A.~G.}\ \bibnamefont {White}},\ and\ \bibinfo {author} {\bibfnamefont {T.~J.}\ \bibnamefont {Weinhold}},\ }\bibfield  {title} {\bibinfo {title} {Highly efficient storage of cavity {{SPDC}} single photons in room temperature gradient echo memory},\ }\href {https://doi.org/10.1063/5.0207712} {\bibfield  {journal} {\bibinfo  {journal} {APL Quantum}\ }\textbf {\bibinfo {volume} {1}},\ \bibinfo {pages} {036102} (\bibinfo {year} {2024})}\BibitemShut {NoStop}%
\bibitem [{\citenamefont {Tirole}\ \emph {et~al.}(2023)\citenamefont {Tirole}, \citenamefont {Vezzoli}, \citenamefont {Galiffi}, \citenamefont {Robertson}, \citenamefont {Maurice}, \citenamefont {Tilmann}, \citenamefont {Maier}, \citenamefont {Pendry},\ and\ \citenamefont {Sapienza}}]{tiroleDoubleslitTimeDiffraction2023}%
  \BibitemOpen
  \bibfield  {author} {\bibinfo {author} {\bibfnamefont {R.}~\bibnamefont {Tirole}}, \bibinfo {author} {\bibfnamefont {S.}~\bibnamefont {Vezzoli}}, \bibinfo {author} {\bibfnamefont {E.}~\bibnamefont {Galiffi}}, \bibinfo {author} {\bibfnamefont {I.}~\bibnamefont {Robertson}}, \bibinfo {author} {\bibfnamefont {D.}~\bibnamefont {Maurice}}, \bibinfo {author} {\bibfnamefont {B.}~\bibnamefont {Tilmann}}, \bibinfo {author} {\bibfnamefont {S.~A.}\ \bibnamefont {Maier}}, \bibinfo {author} {\bibfnamefont {J.~B.}\ \bibnamefont {Pendry}},\ and\ \bibinfo {author} {\bibfnamefont {R.}~\bibnamefont {Sapienza}},\ }\bibfield  {title} {\bibinfo {title} {Double-slit time diffraction at optical frequencies},\ }\href {https://doi.org/10.1038/s41567-023-01993-w} {\bibfield  {journal} {\bibinfo  {journal} {Nature Physics}\ }\textbf {\bibinfo {volume} {19}},\ \bibinfo {pages} {999} (\bibinfo {year} {2023})}\BibitemShut {NoStop}%
\bibitem [{\citenamefont {Hong}\ \emph {et~al.}(2023)\citenamefont {Hong}, \citenamefont {Chen},\ and\ \citenamefont {Chen}}]{hongDelayedchoiceQuantumErasure2023}%
  \BibitemOpen
  \bibfield  {author} {\bibinfo {author} {\bibfnamefont {L.}~\bibnamefont {Hong}}, \bibinfo {author} {\bibfnamefont {Y.}~\bibnamefont {Chen}},\ and\ \bibinfo {author} {\bibfnamefont {L.}~\bibnamefont {Chen}},\ }\bibfield  {title} {\bibinfo {title} {Delayed-choice quantum erasure with nonlocal temporal double-slit interference},\ }\href {https://doi.org/10.1088/1367-2630/acd01f} {\bibfield  {journal} {\bibinfo  {journal} {New Journal of Physics}\ }\textbf {\bibinfo {volume} {25}},\ \bibinfo {pages} {053014} (\bibinfo {year} {2023})}\BibitemShut {NoStop}%
\bibitem [{\citenamefont {Lindner}\ \emph {et~al.}(2005)\citenamefont {Lindner}, \citenamefont {Sch{\"a}tzel}, \citenamefont {Walther}, \citenamefont {Baltu{\v s}ka}, \citenamefont {Goulielmakis}, \citenamefont {Krausz}, \citenamefont {Milo{\v s}evi{\'c}}, \citenamefont {Bauer}, \citenamefont {Becker},\ and\ \citenamefont {Paulus}}]{lindnerAttosecondDoubleSlitExperiment2005}%
  \BibitemOpen
  \bibfield  {author} {\bibinfo {author} {\bibfnamefont {F.}~\bibnamefont {Lindner}}, \bibinfo {author} {\bibfnamefont {M.~G.}\ \bibnamefont {Sch{\"a}tzel}}, \bibinfo {author} {\bibfnamefont {H.}~\bibnamefont {Walther}}, \bibinfo {author} {\bibfnamefont {A.}~\bibnamefont {Baltu{\v s}ka}}, \bibinfo {author} {\bibfnamefont {E.}~\bibnamefont {Goulielmakis}}, \bibinfo {author} {\bibfnamefont {F.}~\bibnamefont {Krausz}}, \bibinfo {author} {\bibfnamefont {D.~B.}\ \bibnamefont {Milo{\v s}evi{\'c}}}, \bibinfo {author} {\bibfnamefont {D.}~\bibnamefont {Bauer}}, \bibinfo {author} {\bibfnamefont {W.}~\bibnamefont {Becker}},\ and\ \bibinfo {author} {\bibfnamefont {G.~G.}\ \bibnamefont {Paulus}},\ }\bibfield  {title} {\bibinfo {title} {Attosecond {{Double-Slit Experiment}}},\ }\href {https://doi.org/10.1103/PhysRevLett.95.040401} {\bibfield  {journal} {\bibinfo  {journal} {Physical Review Letters}\ }\textbf {\bibinfo {volume} {95}},\ \bibinfo {pages} {040401} (\bibinfo {year} {2005})}\BibitemShut {NoStop}%
\bibitem [{\citenamefont {Richter}\ \emph {et~al.}(2015)\citenamefont {Richter}, \citenamefont {Kunitski}, \citenamefont {Sch{\"o}ffler}, \citenamefont {Jahnke}, \citenamefont {Schmidt}, \citenamefont {Li}, \citenamefont {Liu},\ and\ \citenamefont {D{\"o}rner}}]{richterStreakingTemporalDoubleSlit2015}%
  \BibitemOpen
  \bibfield  {author} {\bibinfo {author} {\bibfnamefont {M.}~\bibnamefont {Richter}}, \bibinfo {author} {\bibfnamefont {M.}~\bibnamefont {Kunitski}}, \bibinfo {author} {\bibfnamefont {M.}~\bibnamefont {Sch{\"o}ffler}}, \bibinfo {author} {\bibfnamefont {T.}~\bibnamefont {Jahnke}}, \bibinfo {author} {\bibfnamefont {L.~P.~H.}\ \bibnamefont {Schmidt}}, \bibinfo {author} {\bibfnamefont {M.}~\bibnamefont {Li}}, \bibinfo {author} {\bibfnamefont {Y.}~\bibnamefont {Liu}},\ and\ \bibinfo {author} {\bibfnamefont {R.}~\bibnamefont {D{\"o}rner}},\ }\bibfield  {title} {\bibinfo {title} {Streaking {{Temporal Double-Slit Interference}} by an {{Orthogonal Two-Color Laser Field}}},\ }\href {https://doi.org/10.1103/PhysRevLett.114.143001} {\bibfield  {journal} {\bibinfo  {journal} {Physical Review Letters}\ }\textbf {\bibinfo {volume} {114}},\ \bibinfo {pages} {143001} (\bibinfo {year} {2015})}\BibitemShut {NoStop}%
\bibitem [{\citenamefont {Wollenhaupt}\ \emph {et~al.}(2002)\citenamefont {Wollenhaupt}, \citenamefont {Assion}, \citenamefont {Liese}, \citenamefont {Sarpe-Tudoran}, \citenamefont {Baumert}, \citenamefont {Zamith}, \citenamefont {Bouchene}, \citenamefont {Girard}, \citenamefont {Flettner}, \citenamefont {Weichmann},\ and\ \citenamefont {Gerber}}]{PhysRevLett.89.173001}%
  \BibitemOpen
  \bibfield  {author} {\bibinfo {author} {\bibfnamefont {M.}~\bibnamefont {Wollenhaupt}}, \bibinfo {author} {\bibfnamefont {A.}~\bibnamefont {Assion}}, \bibinfo {author} {\bibfnamefont {D.}~\bibnamefont {Liese}}, \bibinfo {author} {\bibfnamefont {C.}~\bibnamefont {Sarpe-Tudoran}}, \bibinfo {author} {\bibfnamefont {T.}~\bibnamefont {Baumert}}, \bibinfo {author} {\bibfnamefont {S.}~\bibnamefont {Zamith}}, \bibinfo {author} {\bibfnamefont {M.~A.}\ \bibnamefont {Bouchene}}, \bibinfo {author} {\bibfnamefont {B.}~\bibnamefont {Girard}}, \bibinfo {author} {\bibfnamefont {A.}~\bibnamefont {Flettner}}, \bibinfo {author} {\bibfnamefont {U.}~\bibnamefont {Weichmann}},\ and\ \bibinfo {author} {\bibfnamefont {G.}~\bibnamefont {Gerber}},\ }\bibfield  {title} {\bibinfo {title} {Interferences of ultrashort free electron wave packets},\ }\href {https://doi.org/10.1103/PhysRevLett.89.173001} {\bibfield  {journal} {\bibinfo  {journal} {Phys. Rev. Lett.}\ }\textbf {\bibinfo {volume} {89}},\ \bibinfo {pages} {173001} (\bibinfo
  {year} {2002})}\BibitemShut {NoStop}%
\bibitem [{\citenamefont {Szriftgiser}\ \emph {et~al.}(1996)\citenamefont {Szriftgiser}, \citenamefont {Gu\'ery-Odelin}, \citenamefont {Arndt},\ and\ \citenamefont {Dalibard}}]{PhysRevLett.77.4}%
  \BibitemOpen
  \bibfield  {author} {\bibinfo {author} {\bibfnamefont {P.}~\bibnamefont {Szriftgiser}}, \bibinfo {author} {\bibfnamefont {D.}~\bibnamefont {Gu\'ery-Odelin}}, \bibinfo {author} {\bibfnamefont {M.}~\bibnamefont {Arndt}},\ and\ \bibinfo {author} {\bibfnamefont {J.}~\bibnamefont {Dalibard}},\ }\bibfield  {title} {\bibinfo {title} {Atomic wave diffraction and interference using temporal slits},\ }\href {https://doi.org/10.1103/PhysRevLett.77.4} {\bibfield  {journal} {\bibinfo  {journal} {Phys. Rev. Lett.}\ }\textbf {\bibinfo {volume} {77}},\ \bibinfo {pages} {4} (\bibinfo {year} {1996})}\BibitemShut {NoStop}%
\bibitem [{\citenamefont {Gorshkov}\ \emph {et~al.}(2007)\citenamefont {Gorshkov}, \citenamefont {Andr{\'e}}, \citenamefont {Lukin},\ and\ \citenamefont {S{\o}rensen}}]{gorshkovPhotonStorageType2007a}%
  \BibitemOpen
  \bibfield  {author} {\bibinfo {author} {\bibfnamefont {A.~V.}\ \bibnamefont {Gorshkov}}, \bibinfo {author} {\bibfnamefont {A.}~\bibnamefont {Andr{\'e}}}, \bibinfo {author} {\bibfnamefont {M.~D.}\ \bibnamefont {Lukin}},\ and\ \bibinfo {author} {\bibfnamefont {A.~S.}\ \bibnamefont {S{\o}rensen}},\ }\bibfield  {title} {\bibinfo {title} {Photon storage in {{$\Lambda$}} -type optically dense atomic media. {{II}}. {{Free-space}} model},\ }\bibfield  {journal} {\bibinfo  {journal} {Physical Review A}\ }\textbf {\bibinfo {volume} {76}},\ \href {https://doi.org/10.1103/PhysRevA.76.033805} {10.1103/PhysRevA.76.033805} (\bibinfo {year} {2007})\BibitemShut {NoStop}%
\bibitem [{\citenamefont {Everett}\ \emph {et~al.}(2017)\citenamefont {Everett}, \citenamefont {Campbell}, \citenamefont {Cho}, \citenamefont {{Vernaz-Gris}}, \citenamefont {Higginbottom}, \citenamefont {Pinel}, \citenamefont {Robins}, \citenamefont {Lam},\ and\ \citenamefont {Buchler}}]{everettDynamicalObservationsSelfstabilizing2017}%
  \BibitemOpen
  \bibfield  {author} {\bibinfo {author} {\bibfnamefont {J.~L.}\ \bibnamefont {Everett}}, \bibinfo {author} {\bibfnamefont {G.~T.}\ \bibnamefont {Campbell}}, \bibinfo {author} {\bibfnamefont {Y.-W.}\ \bibnamefont {Cho}}, \bibinfo {author} {\bibfnamefont {P.}~\bibnamefont {{Vernaz-Gris}}}, \bibinfo {author} {\bibfnamefont {D.~B.}\ \bibnamefont {Higginbottom}}, \bibinfo {author} {\bibfnamefont {O.}~\bibnamefont {Pinel}}, \bibinfo {author} {\bibfnamefont {N.~P.}\ \bibnamefont {Robins}}, \bibinfo {author} {\bibfnamefont {P.~K.}\ \bibnamefont {Lam}},\ and\ \bibinfo {author} {\bibfnamefont {B.~C.}\ \bibnamefont {Buchler}},\ }\bibfield  {title} {\bibinfo {title} {Dynamical observations of self-stabilizing stationary light},\ }\href {https://doi.org/10.1038/nphys3901} {\bibfield  {journal} {\bibinfo  {journal} {Nature Physics}\ }\textbf {\bibinfo {volume} {13}},\ \bibinfo {pages} {68} (\bibinfo {year} {2017})}\BibitemShut {NoStop}%
\bibitem [{\citenamefont {Campbell}\ \emph {et~al.}(2017)\citenamefont {Campbell}, \citenamefont {Cho}, \citenamefont {Su}, \citenamefont {Everett}, \citenamefont {Robins}, \citenamefont {Lam},\ and\ \citenamefont {Buchler}}]{campbellDirectImagingSlow2017}%
  \BibitemOpen
  \bibfield  {author} {\bibinfo {author} {\bibfnamefont {G.~T.}\ \bibnamefont {Campbell}}, \bibinfo {author} {\bibfnamefont {Y.-W.}\ \bibnamefont {Cho}}, \bibinfo {author} {\bibfnamefont {J.}~\bibnamefont {Su}}, \bibinfo {author} {\bibfnamefont {J.}~\bibnamefont {Everett}}, \bibinfo {author} {\bibfnamefont {N.}~\bibnamefont {Robins}}, \bibinfo {author} {\bibfnamefont {P.~K.}\ \bibnamefont {Lam}},\ and\ \bibinfo {author} {\bibfnamefont {B.}~\bibnamefont {Buchler}},\ }\bibfield  {title} {\bibinfo {title} {Direct imaging of slow, stored and stationary {{EIT}} polaritons},\ }\href {https://doi.org/10.1088/2058-9565/aa7821} {\bibfield  {journal} {\bibinfo  {journal} {Quantum Science and Technology}\ }\textbf {\bibinfo {volume} {2}},\ \bibinfo {pages} {034010} (\bibinfo {year} {2017})}\BibitemShut {NoStop}%
\bibitem [{\citenamefont {{Vernaz-Gris}}\ \emph {et~al.}(2018{\natexlab{b}})\citenamefont {{Vernaz-Gris}}, \citenamefont {Tranter}, \citenamefont {Everett}, \citenamefont {Leung}, \citenamefont {Paul}, \citenamefont {Campbell}, \citenamefont {Lam},\ and\ \citenamefont {Buchler}}]{vernaz-grisHighperformanceRamanMemory2018b}%
  \BibitemOpen
  \bibfield  {author} {\bibinfo {author} {\bibfnamefont {P.}~\bibnamefont {{Vernaz-Gris}}}, \bibinfo {author} {\bibfnamefont {A.~D.}\ \bibnamefont {Tranter}}, \bibinfo {author} {\bibfnamefont {J.~L.}\ \bibnamefont {Everett}}, \bibinfo {author} {\bibfnamefont {A.~C.}\ \bibnamefont {Leung}}, \bibinfo {author} {\bibfnamefont {K.~V.}\ \bibnamefont {Paul}}, \bibinfo {author} {\bibfnamefont {G.~T.}\ \bibnamefont {Campbell}}, \bibinfo {author} {\bibfnamefont {P.~K.}\ \bibnamefont {Lam}},\ and\ \bibinfo {author} {\bibfnamefont {B.~C.}\ \bibnamefont {Buchler}},\ }\bibfield  {title} {\bibinfo {title} {High-performance {{Raman}} memory with spatio-temporal reversal},\ }\href {https://doi.org/10.1364/OE.26.012424} {\bibfield  {journal} {\bibinfo  {journal} {Optics Express}\ }\textbf {\bibinfo {volume} {26}},\ \bibinfo {pages} {12424} (\bibinfo {year} {2018}{\natexlab{b}})}\BibitemShut {NoStop}%
\bibitem [{\citenamefont {Everett}\ \emph {et~al.}(2019)\citenamefont {Everett}, \citenamefont {Higginbottom}, \citenamefont {Campbell}, \citenamefont {Lam},\ and\ \citenamefont {Buchler}}]{everettStationaryLightAtomic2019}%
  \BibitemOpen
  \bibfield  {author} {\bibinfo {author} {\bibfnamefont {J.~L.}\ \bibnamefont {Everett}}, \bibinfo {author} {\bibfnamefont {D.~B.}\ \bibnamefont {Higginbottom}}, \bibinfo {author} {\bibfnamefont {G.~T.}\ \bibnamefont {Campbell}}, \bibinfo {author} {\bibfnamefont {P.~K.}\ \bibnamefont {Lam}},\ and\ \bibinfo {author} {\bibfnamefont {B.~C.}\ \bibnamefont {Buchler}},\ }\bibfield  {title} {\bibinfo {title} {Stationary {{Light}} in {{Atomic Media}}},\ }\href {https://doi.org/10.1002/qute.201800100} {\bibfield  {journal} {\bibinfo  {journal} {Advanced Quantum Technologies}\ }\textbf {\bibinfo {volume} {2}},\ \bibinfo {pages} {1800100} (\bibinfo {year} {2019})}\BibitemShut {NoStop}%
\bibitem [{\citenamefont {Dennis}\ \emph {et~al.}(2013)\citenamefont {Dennis}, \citenamefont {Hope},\ and\ \citenamefont {Johnsson}}]{dennisXMDS2FastScalable2013}%
  \BibitemOpen
  \bibfield  {author} {\bibinfo {author} {\bibfnamefont {G.~R.}\ \bibnamefont {Dennis}}, \bibinfo {author} {\bibfnamefont {J.~J.}\ \bibnamefont {Hope}},\ and\ \bibinfo {author} {\bibfnamefont {M.~T.}\ \bibnamefont {Johnsson}},\ }\bibfield  {title} {\bibinfo {title} {{{XMDS2}}: {{Fast}}, scalable simulation of coupled stochastic partial differential equations},\ }\href {https://doi.org/10.1016/j.cpc.2012.08.016} {\bibfield  {journal} {\bibinfo  {journal} {Computer Physics Communications}\ }\textbf {\bibinfo {volume} {184}},\ \bibinfo {pages} {201} (\bibinfo {year} {2013})}\BibitemShut {NoStop}%
\bibitem [{\citenamefont {{Pellat-Finet}}(1994)}]{pellat-finetFresnelDiffractionFractionalorder1994}%
  \BibitemOpen
  \bibfield  {author} {\bibinfo {author} {\bibfnamefont {P.}~\bibnamefont {{Pellat-Finet}}},\ }\bibfield  {title} {\bibinfo {title} {Fresnel diffraction and the fractional-order {{Fourier}} transform},\ }\href {https://doi.org/10.1364/OL.19.001388} {\bibfield  {journal} {\bibinfo  {journal} {Optics Letters}\ }\textbf {\bibinfo {volume} {19}},\ \bibinfo {pages} {1388} (\bibinfo {year} {1994})}\BibitemShut {NoStop}%
\bibitem [{\citenamefont {Kolner}\ and\ \citenamefont {Nazarathy}(1989)}]{kolnerTemporalImagingTime1989}%
  \BibitemOpen
  \bibfield  {author} {\bibinfo {author} {\bibfnamefont {B.~H.}\ \bibnamefont {Kolner}}\ and\ \bibinfo {author} {\bibfnamefont {M.}~\bibnamefont {Nazarathy}},\ }\bibfield  {title} {\bibinfo {title} {Temporal imaging with a time lens},\ }\href {https://doi.org/10.1364/OL.14.000630} {\bibfield  {journal} {\bibinfo  {journal} {Optics Letters}\ }\textbf {\bibinfo {volume} {14}},\ \bibinfo {pages} {630} (\bibinfo {year} {1989})}\BibitemShut {NoStop}%
\bibitem [{\citenamefont {Shah}\ and\ \citenamefont {Fan}(2021)}]{shahFrequencySuperresolutionSpectrotemporal2021}%
  \BibitemOpen
  \bibfield  {author} {\bibinfo {author} {\bibfnamefont {M.}~\bibnamefont {Shah}}\ and\ \bibinfo {author} {\bibfnamefont {L.}~\bibnamefont {Fan}},\ }\bibfield  {title} {\bibinfo {title} {Frequency {{Superresolution}} with {{Spectrotemporal Shaping}} of {{Photons}}},\ }\href {https://doi.org/10.1103/PhysRevApplied.15.034071} {\bibfield  {journal} {\bibinfo  {journal} {Physical Review Applied}\ }\textbf {\bibinfo {volume} {15}},\ \bibinfo {pages} {034071} (\bibinfo {year} {2021})}\BibitemShut {NoStop}%
\bibitem [{\citenamefont {Mower}\ \emph {et~al.}(2013)\citenamefont {Mower}, \citenamefont {Zhang}, \citenamefont {Desjardins}, \citenamefont {Lee}, \citenamefont {Shapiro},\ and\ \citenamefont {Englund}}]{mowerHighdimensionalQuantumKey2013}%
  \BibitemOpen
  \bibfield  {author} {\bibinfo {author} {\bibfnamefont {J.}~\bibnamefont {Mower}}, \bibinfo {author} {\bibfnamefont {Z.}~\bibnamefont {Zhang}}, \bibinfo {author} {\bibfnamefont {P.}~\bibnamefont {Desjardins}}, \bibinfo {author} {\bibfnamefont {C.}~\bibnamefont {Lee}}, \bibinfo {author} {\bibfnamefont {J.~H.}\ \bibnamefont {Shapiro}},\ and\ \bibinfo {author} {\bibfnamefont {D.}~\bibnamefont {Englund}},\ }\bibfield  {title} {\bibinfo {title} {High-dimensional quantum key distribution using dispersive optics},\ }\href {https://doi.org/10.1103/PhysRevA.87.062322} {\bibfield  {journal} {\bibinfo  {journal} {Physical Review A}\ }\textbf {\bibinfo {volume} {87}},\ \bibinfo {pages} {062322} (\bibinfo {year} {2013})}\BibitemShut {NoStop}%
\bibitem [{\citenamefont {Serino}\ \emph {et~al.}(2025)\citenamefont {Serino}, \citenamefont {Eigner}, \citenamefont {Brecht},\ and\ \citenamefont {Silberhorn}}]{serinoProgrammableTimefrequencyModesorting2025}%
  \BibitemOpen
  \bibfield  {author} {\bibinfo {author} {\bibfnamefont {L.}~\bibnamefont {Serino}}, \bibinfo {author} {\bibfnamefont {C.}~\bibnamefont {Eigner}}, \bibinfo {author} {\bibfnamefont {B.}~\bibnamefont {Brecht}},\ and\ \bibinfo {author} {\bibfnamefont {C.}~\bibnamefont {Silberhorn}},\ }\bibfield  {title} {\bibinfo {title} {Programmable time-frequency mode-sorting of single photons with a multi-output quantum pulse gate},\ }\href {https://doi.org/10.1364/OE.544206} {\bibfield  {journal} {\bibinfo  {journal} {Optics Express}\ }\textbf {\bibinfo {volume} {33}},\ \bibinfo {pages} {5577} (\bibinfo {year} {2025})}\BibitemShut {NoStop}%
\end{thebibliography}%
\end{document}